\documentclass[12pt]{article}
\usepackage{amsmath,graphicx}
\usepackage{hyperref}
\usepackage{caption}
\newcommand{\dietab}{[Dietrich, 2008a,b]}
\newcommand{\oldabstract}[1]{}{}

\title{Positional Effects on Citation and Readership in arXiv}
\author{
Asif-ul Haque\footnote{asif@cs.cornell.edu}, Paul Ginsparg\footnote{ginsparg@cornell.edu}\\
Dept of Information Science, Cornell University, Ithaca, NY 14853\\
}
\date{}

\begin{document}
\maketitle

\begin{abstract}
\baselineskip15pt minus 3pt 

arXiv.org mediates contact with the literature for entire scholarly communities,
both through provision of archival access and through daily email and web announcements
of new materials, potentially many screenlengths long.
We confirm and extend a surprising correlation between article position
in these initial announcements, ordered by submission time,
and later citation impact, due primarily to intentional ``self-promotion" on the part of authors.
A pure ``visibility" effect was also present: the subset of articles accidentally in early positions
fared measurably better in the long-term citation record than those lower down.
Astrophysics articles announced in position 1, for example, overall received a
median number of citations 83\% higher, while those there accidentally had a 44\% visibility boost.
For two large subcommunities of theoretical high energy physics,
hep-th and hep-ph articles announced in position 1 had median numbers of citations 50\%
and 100\% larger than for positions 5--15, and the subsets
there accidentally had visibility boosts of 38\% and 71\%.

We also consider the positional effects on early readership.
The median numbers of early full text downloads for astro-ph, hep-th,
and hep-ph articles announced in position 1 were 82\%, 61\%, and 58\%
higher than for lower positions, respectively, and those there accidentally
had medians visibility-boosted by 53\%, 44\%, and 46\%.
Finally, we correlate a variety of readership features with long-term citations,
using machine learning methods, thereby extending previous results
on the predictive power of early readership in a broader context.
We conclude with some observations on impact metrics and dangers of
recommender mechanisms.

\end{abstract}

\newpage

\section{Introduction}

\begin{figure}[h!]
\centering
\includegraphics[scale=0.4]{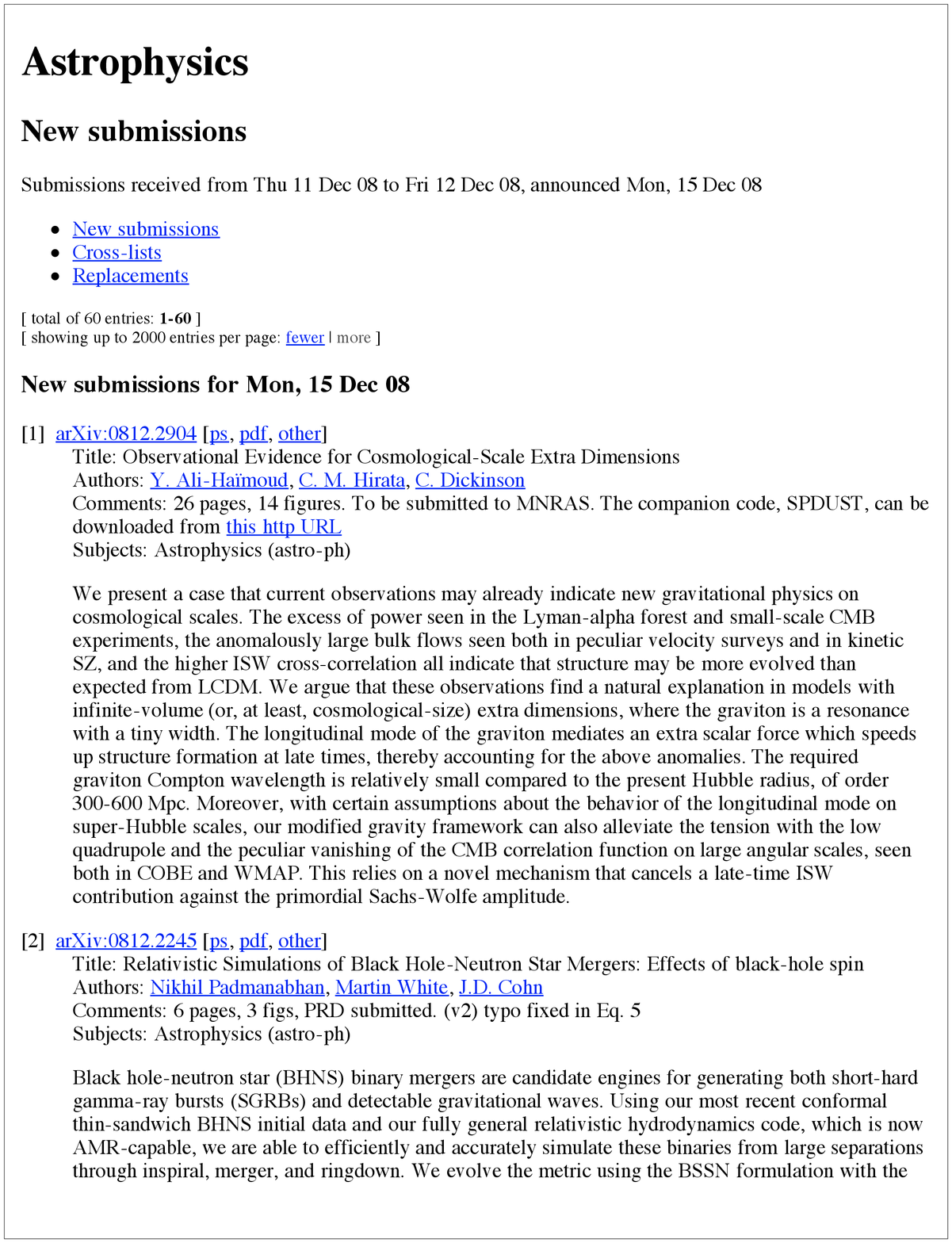}
\caption{\small New astro-ph listings, from http://arXiv.org/list/astro-ph/new . Note that a standard sized Web or e-mail browser window may not accommodate even the full entries in the first two positions without requiring scrolling down. The astro-ph listings averaged roughly thirty such entries every weekday during the period studied here.}
\label{fig:astronew}
\end{figure}

The arXiv\footnote{http://arXiv.org/. For a recent overview, see \cite{Ginsparg07}.} repository currently contains over 500,000 documents and during calendar year 2009 is growing at a rate of over 64,000 new submissions per year. For over a decade, it has been the primary means of access to the research literature in many fields of physics and in some related fields. Its log data provides the basis for many studies of user behavior during this unique transition period from print to electronic medium.
The arXiv corpus  is divided into different subject areas, with corresponding constituent subcommunities. Each of these subcommunities receives notifications each weekday of new articles received in the relevant subject area, either by subscription to email announcements or by checking the web page of newly received submissions in the relevant subject area, updated daily
(or, equivalently, through the associated RSS feed).
These daily listings, viewed either through a web browser or email client, consist of standard metadata, including title and author information, and as well the  full abstracts. As depicted in fig.~\ref{fig:astronew}, this means that it is necessary to scroll down to see beyond the entry in the second position, and to scroll down many times to see the entries in positions near the end of the daily announcements. 
While the overall order of articles is retained when browsing through the archival monthly listings, no trace of the boundaries between days is retained, hence the daily positional information is lost, and of course articles retain no vestige of their position in original daily announcement when retrieved via the search interface.

In what follows here, we investigate the effect of article position in these daily announcements  for certain physics subfields of arXiv, a purely short-term phenomenon, on citations received over the long-term.
This effect for the astro-ph subject area, primarily used by astrophysicists, was first considered in \dietab.
Here we will consider as well two other communities of users, those of the hep-th and hep-ph subject areas (``High Energy Physics -- Theory'' and ``High Energy Physics -- Phenomenology'').
The hep-th subject area is the original arXiv subject area initiated in mid 1991, covering highly theoretical areas of particle physics such as string theory. The hep-ph subject area was started in early 1992, covering areas of theoretical particle physics more directly related to experiment. During the 2002--2004 periods to be studied here, hep-th and hep-ph received an average of roughly 3320 and 4110 new submissions per year, respectively. The astro-ph area, started later in 1992, is an amalgam of many types of relevant theory and experiment, from stellar to galactic to cosmological, and by 2005 had grown to exceed the combined size of the High Energy Physics subject areas.\footnote{http://arxiv.org/Stats/hcamonthly.html}.  The astro-ph subject area averaged roughly 7720 new submissions per year from 2002--2004, and grew to over 9000 new submissions per year in 2005--2006.

A strong correlation between the position of articles in their initial announcement and the number of citations later received
was found in \dietab. Since position in the daily announcement of newly received submissions is a one-day artifact, visible only that day and with no trace afterwards, it is extraordinarily surprising that it could nonetheless be correlated with long-term citation counts, accumulated years later.  Due to the weight given to citations as a measure of research impact, it is important to verify such an unexpected effect by different methods, and assess whether some analog exists as well in other communities.   Our results here confirm the effect discovered in
\dietab, and suggest that arXiv subject area organization and interface design should be reconsidered either to utilize or to counter such unintentional biases.

It is evident to readers that a fraction of authors, working entirely within the established operating procedures for the site, has been jockeying for top position in the daily announcements.
Since late 2001, the policy has been that submissions received until 16:00 US eastern time (EST/EDT) on a given weekday are announced at 20:00 eastern time, and submissions received after that deadline are announced the following day, in rough\footnote{See sec.~\ref{subsec:PW} for important exceptions.} order of receipt.
Articles submitted shortly after 16:00 will thus be listed at or near the top of the next day's announcement, and will potentially receive greater visibility.  Submitters are evidently conforming their schedules to take advantage of some presumed benefit to the greater visibility afforded by submitting within this time window.

\begin{figure}[h]
\centering
\includegraphics[scale=0.7]{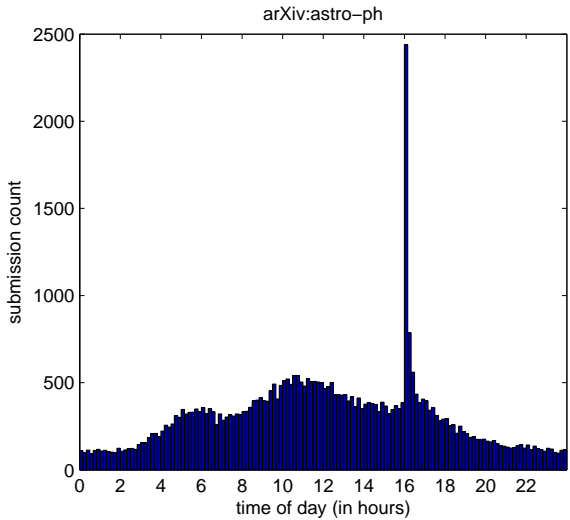}
\caption{\small
Number of astro-ph submissions by time of day, in 10 minute bins, during the period Jan 2002 -- Mar 2007.}
\label{fig:astrosubhist}
\end{figure}

Fig.~\ref{fig:astrosubhist} shows the submission counts, broken down by the time of submission, of arXiv:astro-ph from the beginning of 2002 through the end of Mar 2007. The spike in submissions
corresponds to the period 16:00--16:10. That ten minute bin contains 5 times as many submissions as any other bin outside of the 16:00--16:30 period. Other variations during the day visible in the figure correlate with periodicity of overall activity levels, resulting from the effects of users in different timezones. (The period between 10 a.m.\ and noon eastern time, for example, corresponds to late afternoon in western Europe and early morning in western U.S.
The server itself is not affected by any excessive operating load during the 16:00 period, since the submissions are automatically serialized by time of receipt. Typical submissions take under a second to process, and no noticeable processing queue develops from the [at most] few tens of submissions in that initial minute on a busy day, while the server simultaneously processes multiple retrievals and searches per second.  The average submission rate during the rest of the day is roughly one new submission every six minutes.)

It is important to note that the positional effects are potentially much more dramatic than, say, the corresponding effects in  presentation of search results.  In the latter case, typically ten results are presented on a single web page, with each result entry reduced to a small number of lines of key text.  Eye-tracking studies \cite{Granka04} have shown the extent to which users nonetheless tend to focus only on the top few entries.  In the case of arXiv announcements, on the other hand, the entries consist of entire abstracts (see fig.~\ref{fig:astronew}). Only the first two entries are visible in a standard sized Web or e-mail browser window, and it is necessary to scroll down to see the remainder.  The situation is thus more comparable to viewing successive pages of search results, where for example analysis of log data in  \cite{Fort06} suggested a click probability that decreased with result rank as $r^{-1.63}$.

In the sections below, we consider the positional effects on both citation and readership, in an attempt to understand author and reader behavior, and ascertain whether the policies of the arXiv system itself need modification to counter any unexpected long-term consequences of a seeming short-term artifact.

\section{Effects on Citation}

\subsection{Previous Work}
\label{subsec:PW}

\cite{Dietrich07} used the SPIRES High-Energy Physics Literature Database\footnote{http://www.slac.stanford.edu/spires/hep/}
to reconstruct the daily arXiv astro-ph mailings from Jul 2002 through Dec 2005,
giving the articles at least a year to gather citations. The citations were collected from 
the SAO/NASA Astrophysics Data System (ADS) bibliographic services\footnote{http://adsabs.harvard.edu} in December 2006.
It was inferred that on average, articles  in positions 1 get $89.8\pm9.0$ citations,
while those in positions 10--40 get $44.6\pm0.9$ citations. 
Three possible explanations were suggested for this: self-promotion bias (SP), visibility bias (V),  and geographic bias (G).
The self-promotion argument assumes that authors can intuit in advance the quality of their articles and specifically aim to promote the better ones through early submission.\footnote{This is related in spirit to the `self-selection' postulate \cite{Kurtz05} ,  which suggests that more prestigious articles, i.e., those more likely to be cited, are more likely to be made freely accessible.
In the current context, the suggestion is that those articles are as well promoted by authors to the top of a daily list of new freely accessible articles.}
 Enough of these higher quality articles are submitted in the critical time window to result in the measured citation advantage for submissions in the first few positions.
The visibility  argument is that the initial higher visibility translates to higher readership, and some fraction of that higher readership translates to higher citations later on.  The geographic argument is that articles submitted during the critical period are more likely to come from North America due to timezone differences, and those might be more likely to be cited for other reasons.
Comparing overall citation trajectories of submissions from Europe and North America, however, permitted exclusion of the geographic bias
in \cite{Dietrich07}, and it will not be considered further here.

Using submission times later provided from arXiv log data, a subsquent comparison of three sets of articles was undertaken in \cite{Dietrich08} to disentangle the SP and V biases.
The first set contained articles that appeared in the first three positions and were submitted within the first five minutes after the deadline, hence inferred to have been submitted with an intention to be listed at or near the top.
The second set contained articles that were submitted after the first ninety minutes, and yet appeared in the first three positions.\footnote{This can happen either because there were few or no early submissions, or because an administrative removal of an early article caused a later submitted article to be shifted to that earlier position to fill the gap.} These are assumed not to be self-promoted. The last set contained articles in positions 26--30.
It was observed that the self-promoted articles received more citations than those in the other two sets. The articles that fortuitously appeared near the top, however, also appear to receive more citations than had they appeared in a lower position, indicating as well some visibility bias.  The increase in citations due to the visibility bias was found to be smaller than that due to the self-promotion bias.

The methodology used in \dietab\ to quantify the citation effects involves fitting the citation distributions to a power law, excluding the regimes of data that do not follow the power law (the head and the tail of the distributions), and averaging the rest.
Power law fitting can be tricky \cite{Newman04}, and as described in Appendix~\ref{sec:appPLF}, the above methodology results in inadvertent biases, including using only a portion of the data.  Due to the sociological importance of the result, it is useful to reconsider the results of \dietab\ using slightly different methods.

\subsection{Methodology}

For heavy-tailed distributions such as power laws, the mean can be strongly affected by the large values at the tail. A more robust statistic is the median, which is not affected by the large values, and is also representative of the large number of small values in the sample set. For this reason, nonparametric statistical methods often use the median. More generally, we can consider the $k^\text{th}$ percentile as the aggregate measure of a set of values. If the quartiles (25$^{\rm th}$ and 75$^{\rm th}$ percentiles, usually denoted $Q1$ and $Q3$, respectively) and the median of a distribution are larger than the same quantitites of another distribution (at a statistically significant level), then stochastic dominance (see Appendix~\ref{sec:appPLF}) is likely.  The interquartile range (the difference $Q3-Q1$) measures the spread of the distribution, analogous to the variance of a normal distribution.
 We analyze the citation data by presenting plots of the median and the quartiles and 
 check for statistical significance, using
the nonparametric  Mann-Whitney U (also known as Wilcoxon rank-sum test) and Kolmogorov-Smirnov tests \cite{Gibbons97}.

We consider the 23,165 arXiv astro-ph articles from the beginning of 2002 through the end of 2004, announced in 777 daily announcements (via one-time email announcements and web pages daily updated), with an average mailing containing 29.8 papers. 
The citations were collected from NASA's Astrophysics Data System (ADS) Bibliographic Services in August 2008, giving the articles over three and half years to gather citations.
There are thus 777 articles in each the top positions (and roughly that number in the rest of the positions, at least up to the typical number per announcement).

\begin{figure}[h]
\centering
\includegraphics[scale=0.65]{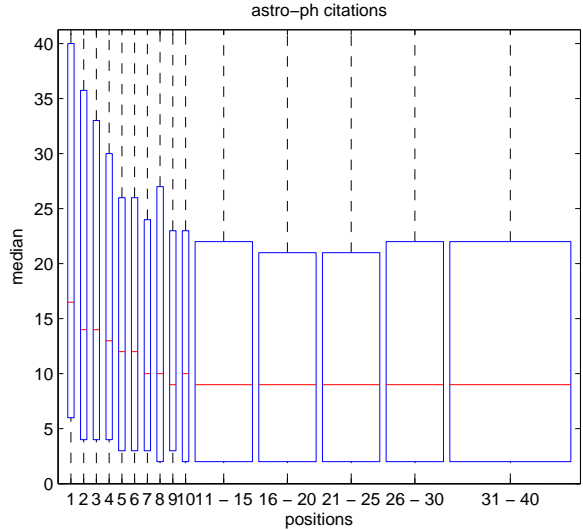}
\caption{\small Box plot of citations for different positions in astro-ph. Boxes represent the interquartile range, bounded above and below by the third and first quartiles, and the red horizontal lines mark the medians.}
\label{fig:astrophcitbox}
\end{figure}

Fig.~\ref{fig:astrophcitbox} shows the median citations and quartiles for each position. The later positions are binned to reduce noise. From position 1, the median decreases until position 5, and beyond position 7 the medians effectively cease changing. The upper quartile (upper boundary of boxes) shows a more pronounced decreasing trend. Even the lower quartiles (lower boundary of boxes) show a decreasing trend.
Statistical significance of these differences is assessed in Appendix~\ref{sec:appSS}.

\subsection{Self-Promotion vs.\ Visibility}
\label{subsec:DSfV}

We now consider the SP and V contributions to increased citations,
taking a different approach from that of \cite{Dietrich08}, as described in sec.~\ref{subsec:PW}.

\begin{figure}[h]
\centering
\includegraphics[scale=0.65]{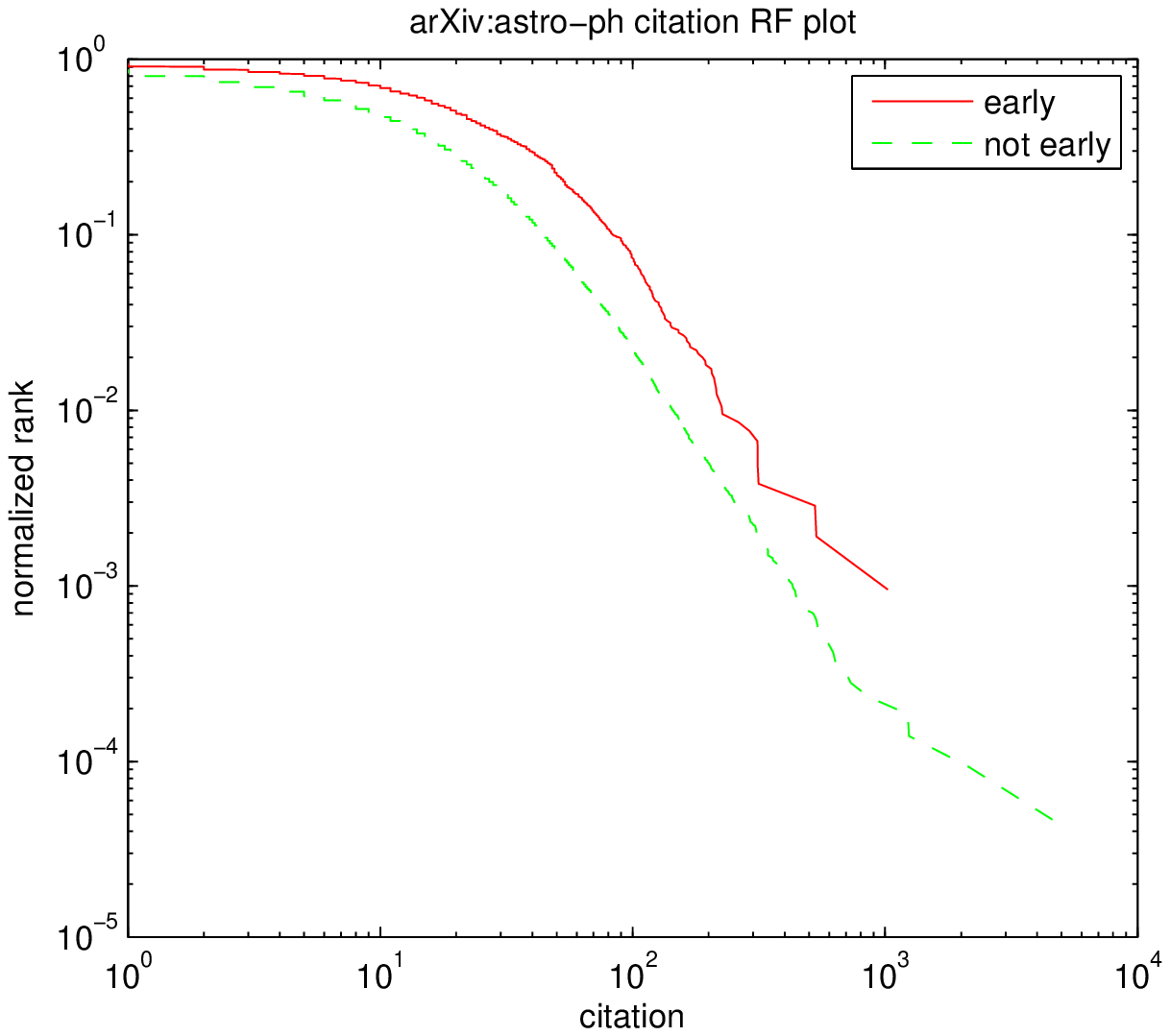}
\caption{\small Rank-Frequency (RF) plot for astro-ph citations. The solid line is for articles submitted within the first 10 minutes after the weekday deadline of 16:00 eastern time. The dashed line is for the articles submitted after the first 30 minutes.}
\label{fig:astrophcitearlyrf}
\end{figure}

 In the astro-ph dataset, we mark those articles submitted in the first 10 minutes after the deadline as ``early'' (E), a time period chosen from fig.~\ref{fig:astrosubhist}. Of the 23,165 articles, 1049 were marked as E, and the vast majority of those are likely to be self-promoted. The articles submitted after the first 30 minutes after deadline were marked ``not early'' (NE). The submitters of these are inferred to be indifferent about the position in the announcements. 643 articles submitted after the first 10 minutes but before the first 30 minutes after deadline were considered as  ambiguous in author intent, so omitted from the analysis (which biases the results in neither direction).

The median citation of the E articles is 20 while that of the NE articles is 9, and the difference in medians is significant using the MWU test at 1\% significance level. The KS test at 1\% significance level shows that the E citation distribution is as well higher than the NE distribution, in the global sense described in Appendix~\ref{sec:appSS}.
The rank--frequency (RF) plots of the two citation distributions, depicted in fig.~\ref{fig:astrophcitearlyrf}, indicate that self-promoting submitters by-and-large do have a good intuition for the likely future impact of their articles.
Not all self-promoted articles, however, receive high citations:  roughly 10\% of the E articles in position 1 have no more than 1 citation.

\begin{table}[h]
\begin{center}
astro-ph:
\begin{tabular}[c]{| c | c | c | c | c | c |}
\hline
Position & 1 & 2 & 3 & 4 & 5 \\
\hline
\hline
Early     & 510 & 289 & 146  & 64  & 24  \\
Not Early & 147 & 299 & 484 & 613 & 694 \\
\hline
\end{tabular}
\caption{\small Number of E and NE articles in arXiv:astro-ph listed at positions 1--5 during the 2002--2004 timeframe.}
\label{table:astrophearlycount}
\end{center}
\end{table}

\begin{figure}[h!]
\centering
\includegraphics[scale=0.65]{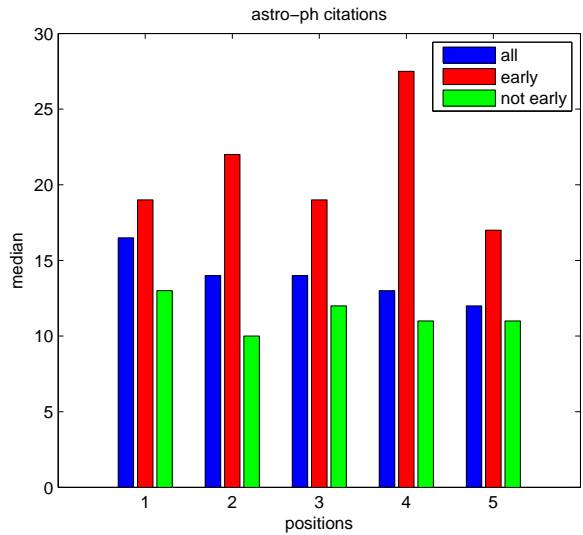}
\caption{\small Median citations for each position for astro-ph announcements from the beginning of 2002 through the end of 2004. The red bars represented the `self-promoted' articles. The non-self-promoted articles in the top few positions, represented by the green bars, nonetheless receive more median citations than those lower down in announcements.}
\label{fig:astrophcitbar}
\end{figure}

To further probe the two biases, we separate out the E articles at each position.
For the top five positions, the numbers of articles are shown in table \ref{table:astrophearlycount}.
Fig.~\ref{fig:astrophcitbar} shows the median number of citations for each position.
The red bars (E articles) characterize the self-promotion effect, while the green bars (NE articles) characterize the visibility effect.
At every position, we see that the effect of self-promotion is much stronger than that of visibility, a difference
significant at 1\% level (MWU test) for the first 4 positions.
The citation advantage of the top few positions is thus largely due to self-promotion, but as we shall see there is as well a visibility effect.

The differences between the blue bars in positions 1 and 5 in fig.~\ref{fig:astrophcitbar} is statistically significant (MWU test at 5\% level), and while it is likely that this difference is entirely due to the SP effect, there is not enough data in the green bars to make a statistically significant statement (at the same 5\% level).  
But we can compare these articles to ones that appeared in lower positions.
The median citation of articles in positions 10--40 is 9, while the median citation of non-SP articles in positions 1-3, i.e., submitted after the first 30 minutes, is 12. 
This difference of 3 citations (significant at the 1\% level) is the extent to which visibility bias contributes to citations.
The non-SP articles are randomly selected, independent of authorship, length, subject area with Astrophysics, or other confounding quality factors, yet solely by virtue of having appeared near the top of a web page or email announcement on one single day, are measured to receive significantly more median citations many years later.\footnote{For comparison with the bins used by \dietab, articles announced in astro-ph positions 1--6, received a median of 14 citations, 55\% higher than 
the median of 9 for those in positions 10--40. The NE articles in positions 1--6 received a median of 11 citations, pure visibility still giving 22\% more median citations than those lower down.}

We have also analyzed the data for the full period in fig.~\ref{fig:astrosubhist}, i.e., the 43,686 astro-ph articles from the beginning from $2002$ till the end of March $2007$ 
announced in 1350 mailings. With citations again collected from ADS in August 2008, this gave at least roughly a year and half for the most recent articles to gather citations.  The resulting graph has the same general form as  fig.~\ref{fig:astrophcitbar}, with greater significance and medians only 10\% to 20\% smaller. But the ``median number of citations'' for the enlarged dataset doesn't correspond to any particular set of articles, because it involves an average over articles of vastly different ages,  with as much as six and a half years to as little as a year and a half to collect citations.  For this reason we used the smaller data set for which the medians do correspond to median numbers of citations for 
4.5--6.5 years old, and don't change appreciably when we further restrict the time window of articles considered. The early timeframe was chosen for stable citation data, although the SP effect became increasingly pronounced in the later data.

\subsection{hep-th and hep-ph}

Having confirmed the self-promotion phenomenon in the astro-ph subject area, we now consider the hep-th and hep-ph subject areas: the largest and most active of arXiv's high energy physics areas.  The 776 daily announcements for those areas during the Jan  2002--Dec 2004 period had averages of 12.8 and 15.9 articles, respectively.

\begin{figure}[h!]
\includegraphics[scale=0.65]{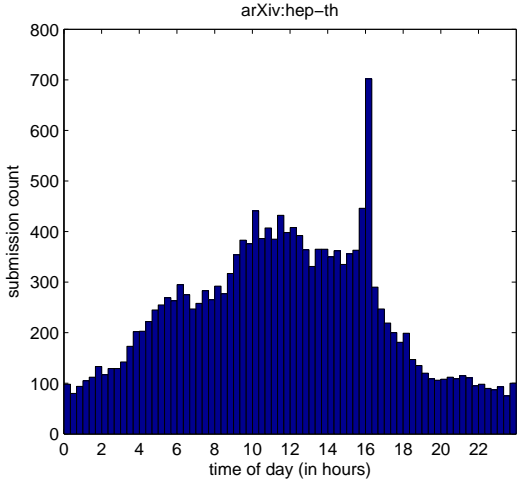}
\quad
\includegraphics[scale=0.65]{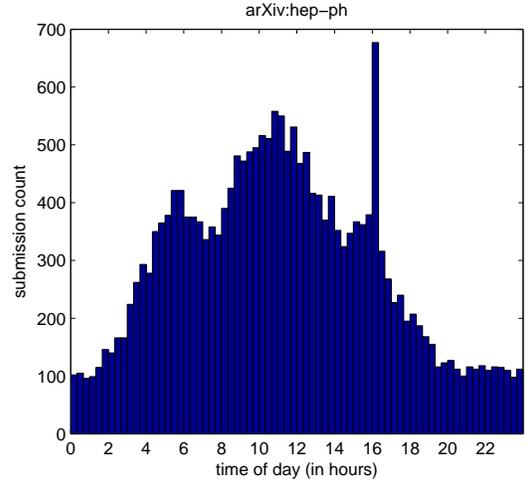}
\centerline{(a)\hskip3in(b)}
\caption{\small Number of (a) hep-th and (b) hep-ph submissions by time of day, in 20 minute bins, during the period Jan 2002 -- Mar 2007.}
\label{fig:hepsubhist}
\end{figure}

Figs.~\ref{fig:hepsubhist}a,b show the number of hep-th and hep-ph submissions
from the beginning of 2002 through Mar 2007, in 20 minute submission bins.
The first 20 minutes after 16:00 eastern time have exceptionally high submission rates, although not as high as astro-ph
(fig.~\ref{fig:astrosubhist}). Articles submitted in this 20 minute period are considered early (E) and the rest are considered not early (NE). We use the articles submitted from Jan 2002 through Dec 2004 for our analysis, for reasons discussed at the end of the previous subsection.
Of the 9,932 total hep-th submissions during this period, 309 were submitted during the first 20 minutes and marked as E;
and of the corresponding 12,281 hep-ph articles, 363 are marked as E, a similar percentage as for hep-th. 
Citations were collected from the SPIRES High-Energy Physics Literature Database in September 2008, giving the articles over three and half years to accumulate citations.  The high energy physics literature, like the astrophysics literature, is served by a relatively small number of conventional published journals, and dominated by a very small number of very large ones.

\begin{figure}[h!]
\includegraphics[scale=0.69]{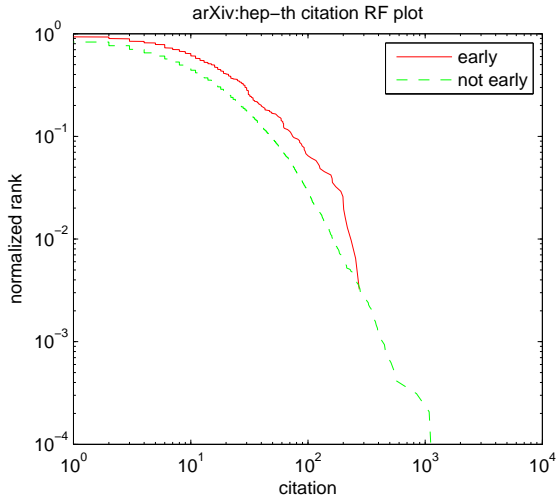}
\quad
\includegraphics[scale=0.69]{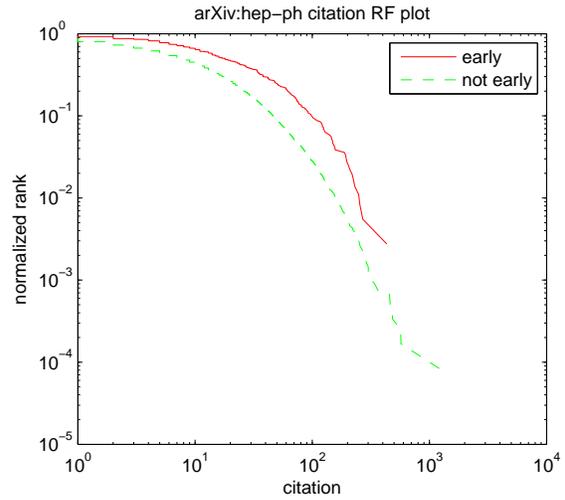}
\centerline{(a)\hskip3in(b)}
\caption{\small Rank-Frequency (RF) plot for (a) hep-th and (b) hep-ph citations. The solid lines represents\ E articles, submitted within the first 20 minutes after the 16:00 eastern time weekday deadline. The dashed lines are for the remaining articles.}
\label{fig:hepcitearlyrf}
\end{figure}

The early hep-th and hep-ph articles are interpreted as self-promoted and, as seen in figs.~\ref{fig:hepcitearlyrf}a,b, their citation distribution stochastically dominates the rest (KS test at 1\% significance level). 
The median citation for hep-th position 1 is 12, while the median citation for positions 4--15 is a significantly lower 8.
Similarly for hep-ph, articles at position 1 have a median citation of 14, while articles at positions 4--15 have a median citation of 7.

\begin{figure}[h!]
\hskip-.2in
\includegraphics[scale=0.60]{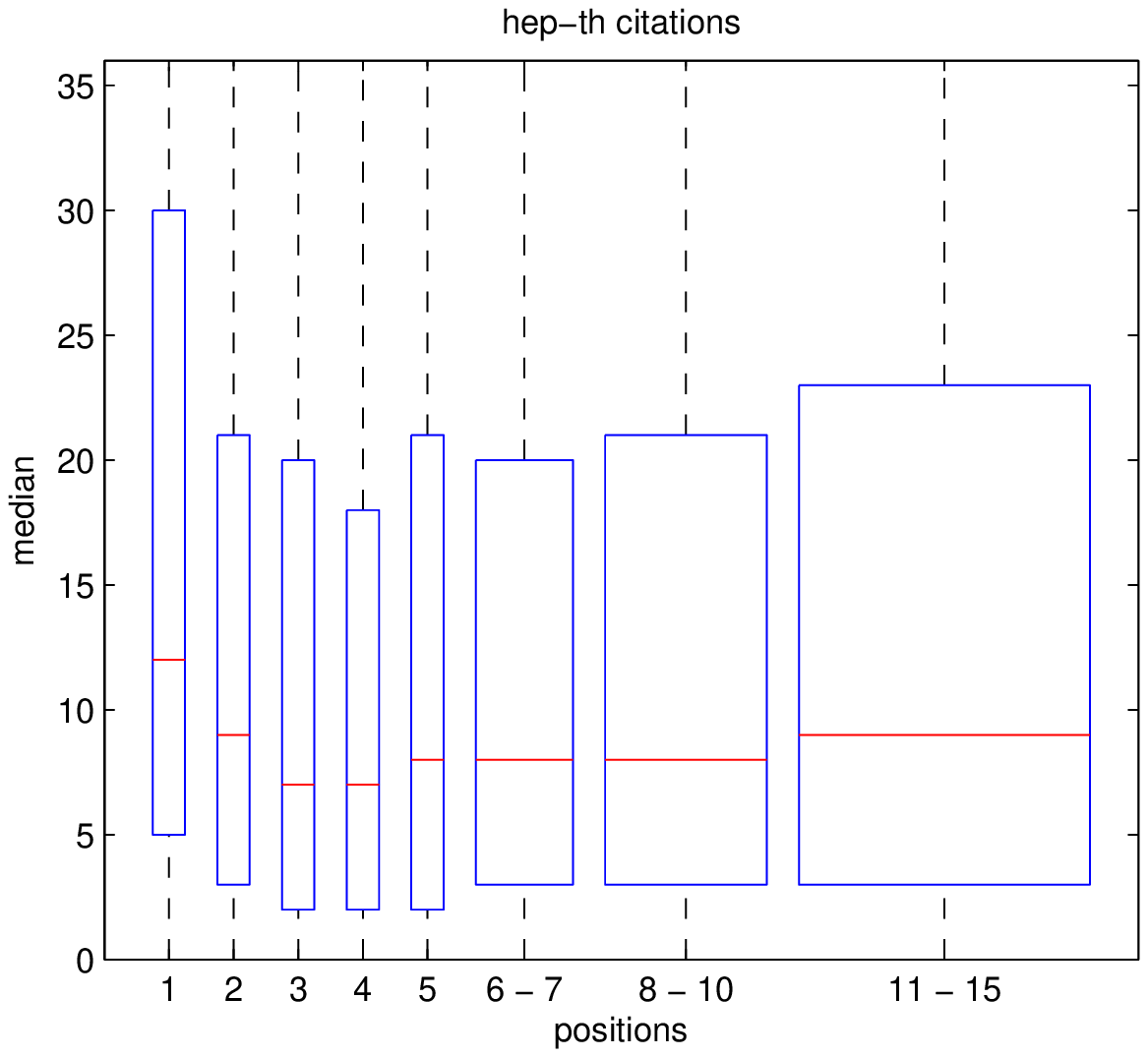}
\quad
\includegraphics[scale=0.60]{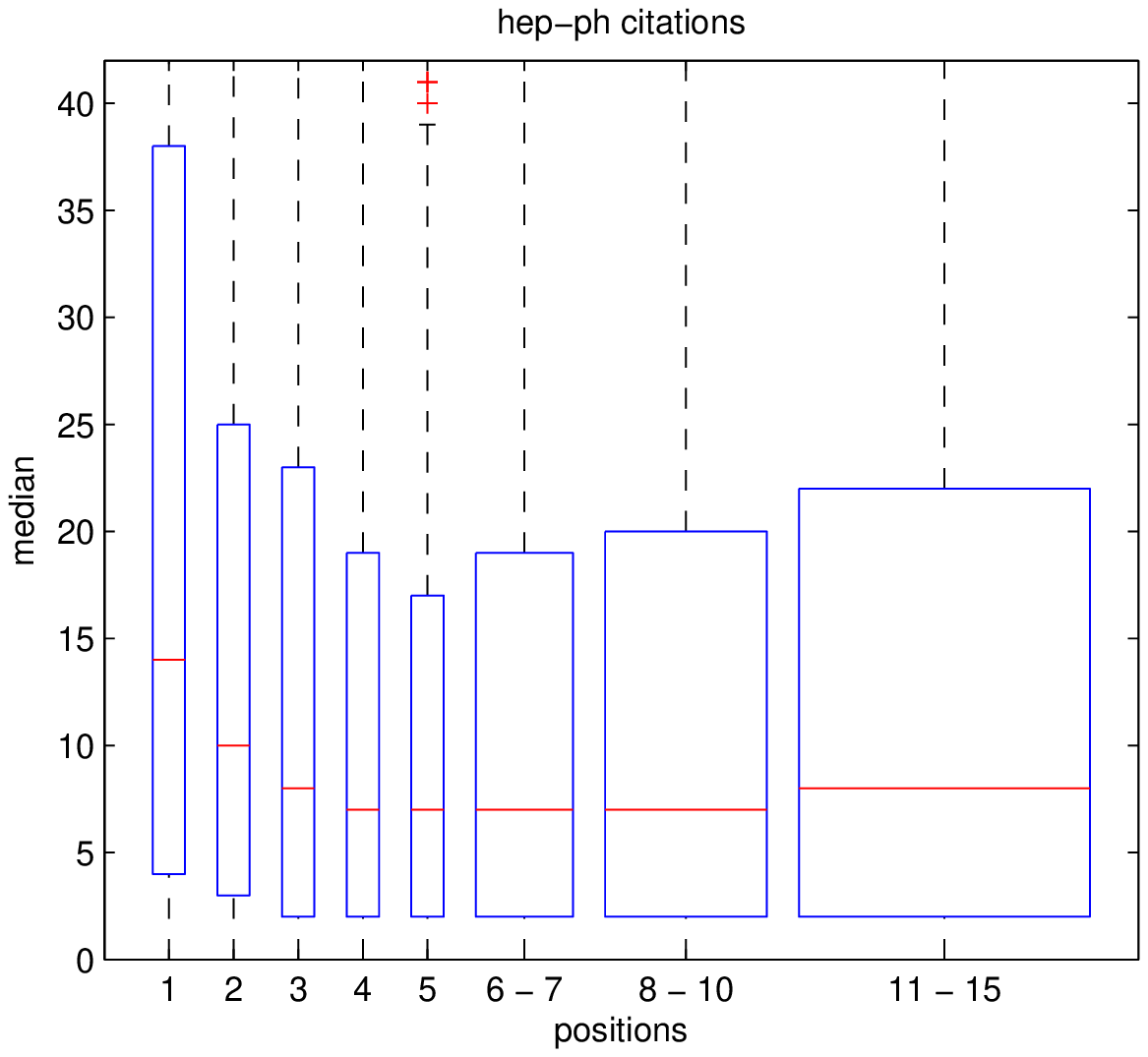}
\centerline{(a)\hskip3in(b)}
\caption{\small Box plot of citations for different positions in (a) hep-th and (b) hep-ph. Boxes depict the interquartile range, and the red lines mark the medians.}
\label{fig:hepcitbox}
\end{figure}

\begin{table}[h!]
hep-th:
\begin{tabular}[c]{| c | c | c | c | c |}
\hline
Position & 1 & 2 & 3 \\
\hline
\hline
Early     & 237 & 58  & 11 \\
Not early & 537 & 715 & 759 \\
\hline
\end{tabular}
\qquad\qquad
hep-ph:
\begin{tabular}[c]{| c | c | c | c | c |}
\hline
Position & 1 & 2 & 3 \\
\hline
\hline
Early     & 282 & 67  & 12 \\
Not early & 492 & 703 & 756 \\
\hline
\end{tabular}
\caption{\small Number of articles in arXiv:hep-th and arXiv:hep-ph listed at positions 1--3  during the 2002--2004 timeframe.}
\label{table:hepearlycount}
\end{table}

\begin{figure}[h!]
\hskip-.2in
\includegraphics[scale=0.63]{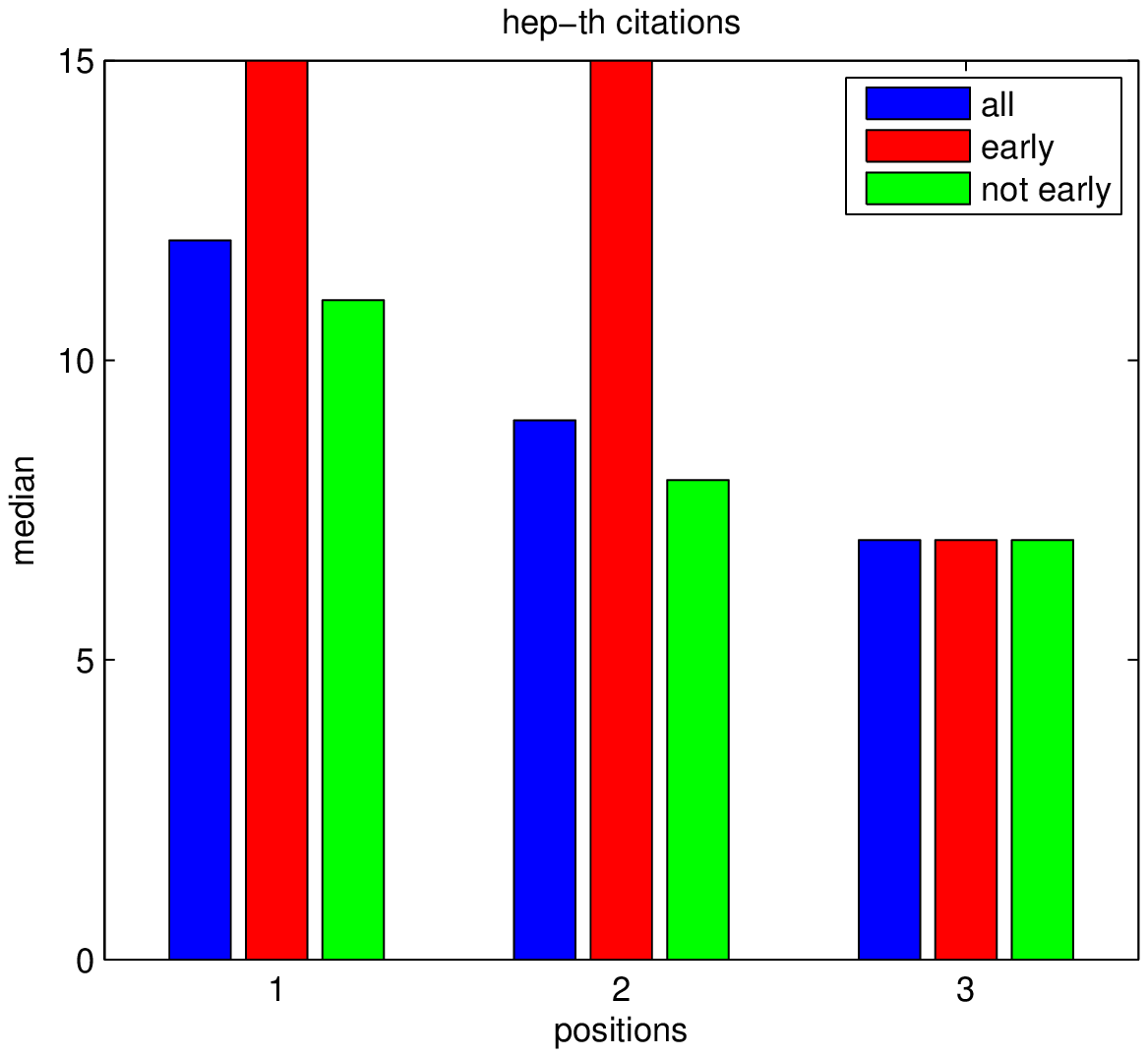}
\quad
\includegraphics[scale=0.63]{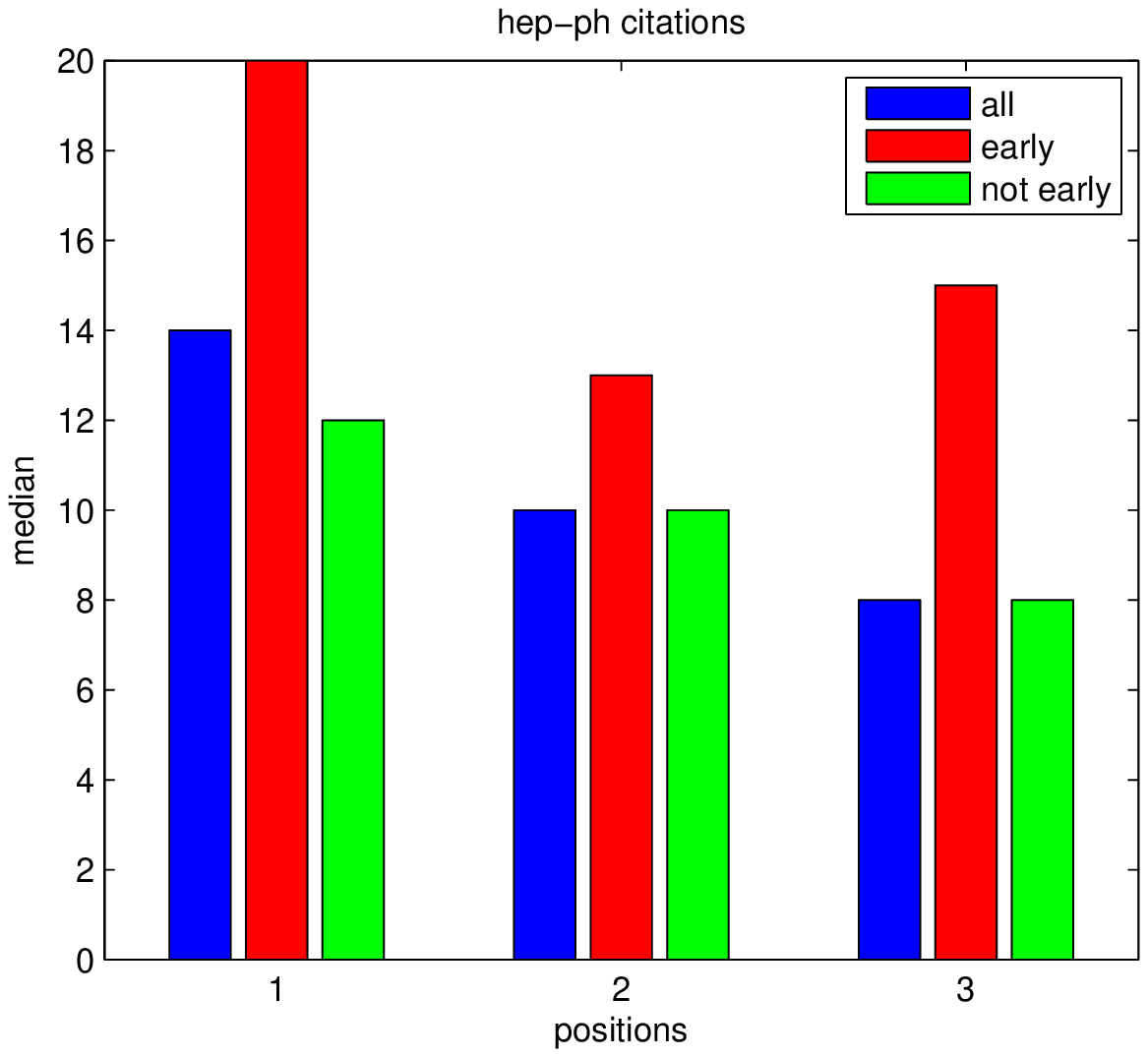}
\centerline{(a)\hskip3in(b)}
\caption{\small Median citations for each position in (a) hep-th and (b) hep-ph, for announcements from the beginning of 2002 through the end of 2004. The red bars represented the `self-promoted' articles.}
\label{fig:hepcitbar}
\end{figure}

Fig.~\ref{fig:hepcitbox} shows the medians and the quartiles of different hep-th and hep-ph positions. The first two positions have median number of citations  significantly higher (at the 1\% level) than the lower positions, and the difference between positions 1 and 2 is particularly striking. 
Fig.~\ref{fig:hepcitbar} disentangles self-promotion and visibility effects and, as in astro-ph,  the self-promotion effect 
(red bars) dominates over the visibility effect (green bars), significant (1\% level) for the first 2 positions. The effect is quite striking for the first position.
The number of articles at each position is shown in table~\ref{table:hepearlycount}. Note that since there were only 11 and 12 early articles  at position 3, respectively, for hep-th and hep-ph, the red bars for this case 
in Figs.~\ref{fig:hepcitbar}a,b are not statistically significant (and similarly for positions 4 and beyond).

\subsubsection*{Visibility}
Although self-promotion is the dominant effect in the positional citation advantage in each of astro-ph, hep-th and hep-ph
(figs.~\ref{fig:astrophcitbar}, \ref{fig:hepcitbar}a,b), there was a pure visibility effect in the astro-ph data and here we find it as well in the hep-th and hep-ph data.
For hep-th, the articles in position 1, but not early (green bar in fig.~\ref{fig:hepcitbar}a), have a median of 11 citations. 
Articles in positions 5--10 have a median of 8 citations. This difference is significant at the 1\% level.
Similarly, for hep-ph, the articles in position 1, but not early (green bar in fig.~\ref{fig:hepcitbar}b), have a significant median citation advantage of $12-7 = 5$ citations over the articles in position 5--10.
The falling trends in the green bars in figs.~\ref{fig:hepcitbar}a,b  capture the beginning of this visibility effect.

\subsection{Discussion}

It is not within the purview of this article to attempt a detailed explanation of why a one-time visibility would leave its trace in the citation record years later.  As we shall see in the readership data in the next section, articles in the top few positions receive more initial downloads, whether or not submitted early (i.e., self-promoted).
The extra initial readership may probabilistically translate into a few early citations, which in turn could cascade into more citations later on.  We could hope to model this in terms of some set of ``fungible'' articles, more or less similar in quality and subject area, with the ones cited determined by something of a social convention, based on artifactual collective effects within the citing community. This would parallel the behavior seen in studies of how social influence affects individual decisions and collective outcome in social networks\cite{Watts2006}. 

Citation practices differ from discipline to discipline, and there there are many known pitfalls of citation as measure of quality. 
Studies of subsets of geoscience \cite{Stewart83}, astrophysics \cite{Baldi98}, and demography \cite{vanDalen2001}
do at least suggest that citations primarily indicate some form of direct intellectual acknowledgement and information flow, rather than primarily reflecting reputational or other secondary social factors.

But other features are known to be correlated to increased citation, including number of authors\footnote{Larger groups could be correlated with more funding and hence better equipment and past track record; see also sec.~\ref{subsec:citread}.}, number of pages, and also specifically visibility factors such as mainstream media coverage or being featured on a journal front cover.
For example, it was shown in \cite{Phillips91} that major media coverage alone could lead to increased citations. Control for other factors in that study was provided by a serendipitous period for which there is a newspaper archive of stories that would have appeared, but were not disseminated due to a distribution strike: the journal research articles that would have been featured in those stories do not exhibit the same citation boost as did articles covered during periods of normal newspaper distribution.
A similar effect can now be expected from visibility in blogspace, or via publicity in either blogspace or the media and amplified through feedback loops between them.  
The analyses of \cite{Dietrich08} and this section were similarly able to isolate the role of visibility by exploiting
the serendipity of randomly selected articles accidentally accorded high visibility without the conscious intent of the authors.

Since a significant component of the citation effect is nonetheless due to intentional self-promotion, it is natural to wonder\footnote{as was indeed wondered by an anonymous referee} whether other forms of additional care taken during the submission process as well correlate with early submission, and hence with more citations in the long term.  For example, it is optional for authors to provide their institutional affiliations parenthetically along with their names in the Author field. We find that 63\% of the early astro-ph submitters provided affiliations, compared to only 43\% of the not early ones.  The total length of the metadata fields in arXiv has always been limited to prevent any one submission from monopolizing too much screen space. (Submissions exceeding the limit are automatically rejected until they are within the limit, just as in this journal's submission process.)  But early submitters nonetheless took maximal advantage within the guidelines: the median length of the title for early submissions was 70 characters, compared to 66 for not early ones (the difference significant at 1\% level KS),  and the median length of abstract was also greater for the earlier submissions, 1177 compared to 1014 characters (i.e., 16 lines compared to 14 in the email announcements, with lines wrapped at the nearest whitespace to under 80 characters per line).

By contrast, early and not early submissions had the same median number of authors (three), the same likelihood of providing initials rather than full first names of authors,
and (reassuringly) there was no tendency for authors of early submissions to have longer last names, 
so the increased length of the overall author field (median of 70 characters compared to 62)
was due entirely to the increased tendency of early submitters to provide author affiliations.
The greater completeness of metadata and inferred submitter effort also correlates with greater citation impact even among only the non self-promoted articles: for not early submission with author affiliations provided, the median number of citations in the 2002--2004 astro-ph dataset was 10 compared to 9 for those without, a statistically significant difference (1\% MWU).
(For the early submissions, the median number of citations was also greater for the submissions that provided affiliation, 20 compared to 19, but the difference was not statistically significant in that case.)

The considerations in this section are also in principle independent of the `citation advantage' sometimes postulated for open access articles, since all of the articles in arXiv are equally open access.  But if the existence of this one-time visibility effect suggests the possibility of an open access advantage, then any analog of the self-promotion effect 
(i.e., that articles more likely to be cited are {\it a priori\/} more likely to be deposited in an open access site)
would have to be eliminated as the underlying cause.

The latter self-selection effect \cite{Kurtz05} was considered further as
a `quality bias' in \cite{Moed07} and \cite{DavFro07}, which studied respectively the citation impact of those articles in the Condensed Matter (cond-mat) and Mathematics (math) sections of arXiv later published in journals, as compared to articles in the same journals but not deposited in arXiv.
Both `early view' (advance availability on arXiv prior to publication in journal) and `quality bias' (higher quality articles more likely to be posted on arXiv) are potential confounding effects that could lead to an artifactual citation advantage, and it was found that correcting for those left no general `open access advantage' for articles deposited in arXiv.  
Similarly, in a study of open access articles published in eleven scientific journals, \cite{Davis08} used a randomized controlled trial to eliminate biases from other quality indicators: whether self-archived, featured front cover of journal, received press-release, and other confounding attributes (nature of article, number of authors and geographic location, number of references, article length, journal impact factor), and later estimated their effect. This study as well found that any citation differences were due to factors other than open access {\it per se\/}:
while those articles randomly assigned open access status received more full text downloads, they were no more likely to be cited a year later.

In Jan 2009, the astro-ph section of arXiv was subdivided into six smaller subsections. It remains possible to receive the combined daily listings for all subsections, but many users expressed a preference to be able to browse only the restricted subsets. This division into smaller announcements will in principle ameliorate some of the positional effects, but not all, since the larger of these subsections still average more than ten new submissions per day.
Some users have suggested randomizing the daily order entirely, either uniformly for everyone, or individually for each user.  Others have pointed out that such a methodology would potentially do a disservice to readers, who may indeed be benefitting from having self-promoted articles brought preferentially to their attention (presuming those really are the more likely to be of importance in the long-run). Perhaps a better methodology is afforded by personalization, by which  users can register to receive daily announcements based on their preferences, and ordered accordingly.\footnote{Such a personalization system has been available to the subset of readers using the myADS features of the NASA ADS system, at http://myads.harvard.edu/ .}
 These preferences can be indicated via a controlled vocabulary of keywords, or via arbitrary search terms, and can be implemented in combination with data from a user's own past on-line reading behavior at the site, on an opt-in basis.

\section{Readership Data}

Since citations can signify some long-term reflection of quality (positive or negative), it is reassuring that the positional advantage of citation is primarily due to self-promotion, rather than to a one-time visibility effect.
In this section, we consider the visibility effect on readership, and more generally consider
how readership features can be used to predict the number of citations of an article.
We will use full-text downloads as a proxy for readership. The download data is from the main arXiv site only, though constitutes a representative sample. It is cleaned of robotic accesses and multiple repeat accesses from the same domain within a small timeframe.  Many articles are made available at the arXiv site in advance of publication by a peer-reviewed journals, though some authors await the results of peer review and make them available at arXiv.org more or less simultaneously with their appearance in a conventional journal.

\subsection{Previous Work}

Past studies have explored the relationship between downloads and citations.
Using ADS data from 7.66 months of 2001, including more than 1.8 million ``reads'', \cite{Kurtz04} studied, among other things, the mean relation between reads and cites, and estimated roughly twenty ADS reads per citation for that period.
\cite{Perneger2004} investigated the relationship between citations and first week's downloads for 153 articles in the British Medical Journal (vol.~318 from 1999), and found that the first week's download activity appeared to capture subsequent article citability.
\cite{Moed2004} computed the correlation between downloads and  citations using a larger sample from the journal Tetrahedron Letters: 1,190 short articles published during the first half of 2001, with about 
410,000 total downloads and 4,300 total citations.
\cite{Brody2005} discussed the correlation between early downloads (minus the first seven days) and citations
of arXiv articles deposited 2000--2002.  The data in this case came only from a  single arXiv mirror, since the more voluminous data from the main site was not publicly available.
\cite{nature08} considered the relation between early downloads (first 90 days) and future citations for  a few hundred articles that appeared in Nature Neuroscience during the period Feb--Dec 2005, and found a usefully predictive correlation, despite a comparatively small level of download activity.

 In what follows here, we use a data set considerably larger than the data used in those studies,
and moreover a different methodology.
Downloads and citations are typical heavy-tailed rather than normal distributions, so measures such as mean and standard deviation are less useful. 
Instead of computing a simple correlation between two variables, we consider the problem as a prediction task and use modern machine learning tools.  Finally, we focus on the positional effect on readership, an effect not considered at all in the above, although any general relation between readership and citation, combined with a positional effect on citations investigated in the previous section, would naturally imply a positional effect on readership.

\subsection{General Pattern}

\begin{figure}[h]
\centering
\includegraphics[scale=0.7]{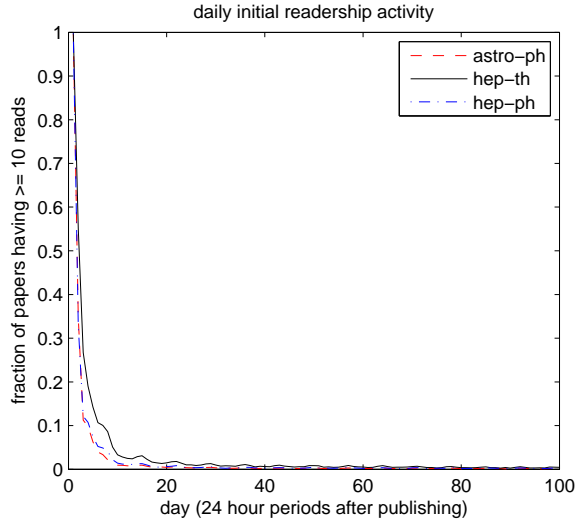}
\caption{\small Fraction (in the subject area) of articles having $\ge 10$ reads on a day.}
\label{fig:initreadact}
\end{figure}

We use the readership for articles in the astro-ph, hep-th and hep-ph subject areas of arXiv received  from Jan 2002 through Mar 2007. The dataset contains the date and time of every full-text download for each article through  the end of 2007.
There is great variation in the temporal readership pattern of articles, 
but the general feature is a burst of initial readership during an  ``active'' period, and only sparse readership thereafter.\footnote{This permits use of the full 5+ years of data, unlike the citation study of the previous section.}
The existence of such an ``active'' period is an indication of the extent to which readers track the research via the daily announcements of new submissions.
In fig.~\ref{fig:initreadact}, we see that almost all articles are downloaded at least 10 times on the day they are first made public, and that fraction then falls rapidly.\footnote{The seven day periodicity in fig.~\ref{fig:initreadact} results from the confluence of  lower weekend readership with announcements of articles being made only on the five weekdays.
We also checked for a possible ``day of the week'' bias, but found that the particular day of the week that an article is announced has no effect on the median number of citations.}
For astro-ph, less than 1\% of the articles have 10 or more downloads per day after the first 10 days. We take 1\% to be the threshold of activity, so the active period for astro-ph is taken to be roughly 10 days.  For hep-th, this period is 25 days, while for hep-ph it is 15 days. The total number of downloads in the active period can be taken as a measure of the initial popularity of an article.

Beyond the active period, typical articles receive no downloads on most days.\footnote{The articles that tend to have the most usage in the long-term are review articles and other pedagogical resources such as lecture notes. Ironically, this long-term usage is frequently not reflected in the citation record. These articles constitute a small enough fraction of the total that they do not skew the data.}
In astro-ph, for example, an average article is downloaded at least once during 12\% of the days of its lifetime. For hep-th and hep-ph, this number is 13\% and 17\% respectively, with a standard deviation of about 10\%.
Readership can therefore be characterized by the number of days an article gets at least some downloads. Since the articles are of varying age in our dataset, we compare their readership activity beyond the active period by using the fraction of days an article gets downloaded at least once.
It is natural to ask if there is a correlation between total initial reads and later (long-term) fraction of days getting some reads.
Table~\ref{table:initreadfracdayscorr} shows that indeed the fraction of later days getting some downloads is quite strongly correlated with initial popularity,
by two common statistical measures.\footnote{
The Pearson correlation coefficient is a parametric statistic computed directly using the values. 
The Spearman correlation coefficient is the nonparametric version of Pearson, replacing the values with their ranks in sorted order. Correlation coefficients range from $-1$ and $+1$, where $+1$ indicates a linear correlation, 0 no correlation, and $-1$ linear anti-correlation.
A value of 0.5 or more is ordinarily considered high.
}

\begin{table}[h]
\centering
\begin{tabular}[c]{| c || c | c | c |}
\hline
 & astro-ph & hep-th & hep-ph \\
\hline
\hline
Pearson & $0.5861$ & $0.7436$ & $0.6625$ \\
Spearman & $0.6716$ & $0.7525$ & $0.6750$ \\
\hline
\end{tabular}
\caption{\small Correlation ($P=0$)
between the number of downloads in the active period with the fraction of days, beyond the active period, an article is downloaded at least once.}
\label{table:initreadfracdayscorr}
\end{table}

\subsection{Positional Effects}
\label{subsec:poseff}

We now examine the relation between article position on the day of announcement and the total number of downloads in the initial active period.

\begin{figure}[h!]
\hskip-15pt
\includegraphics[scale=0.65]{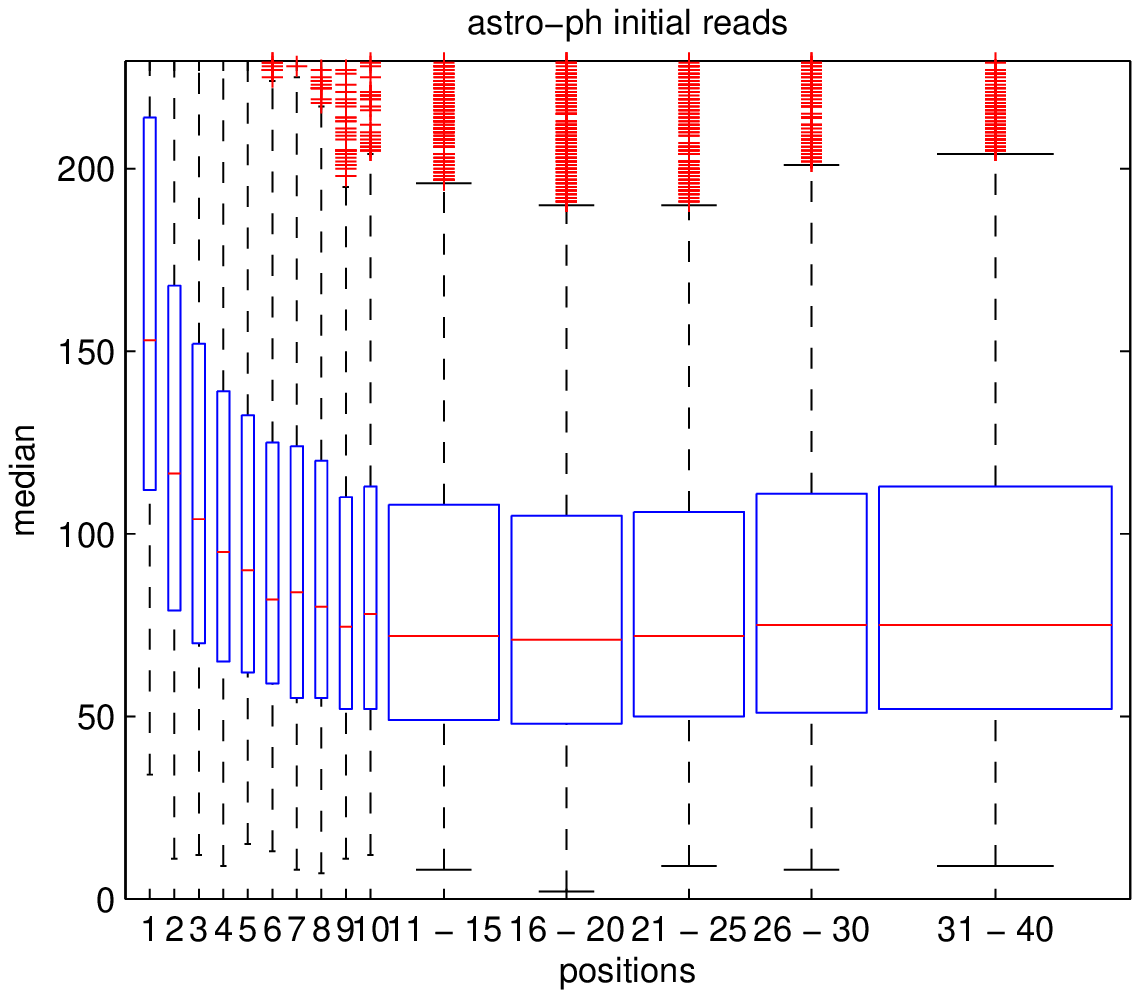}
\quad
\includegraphics[scale=0.65]{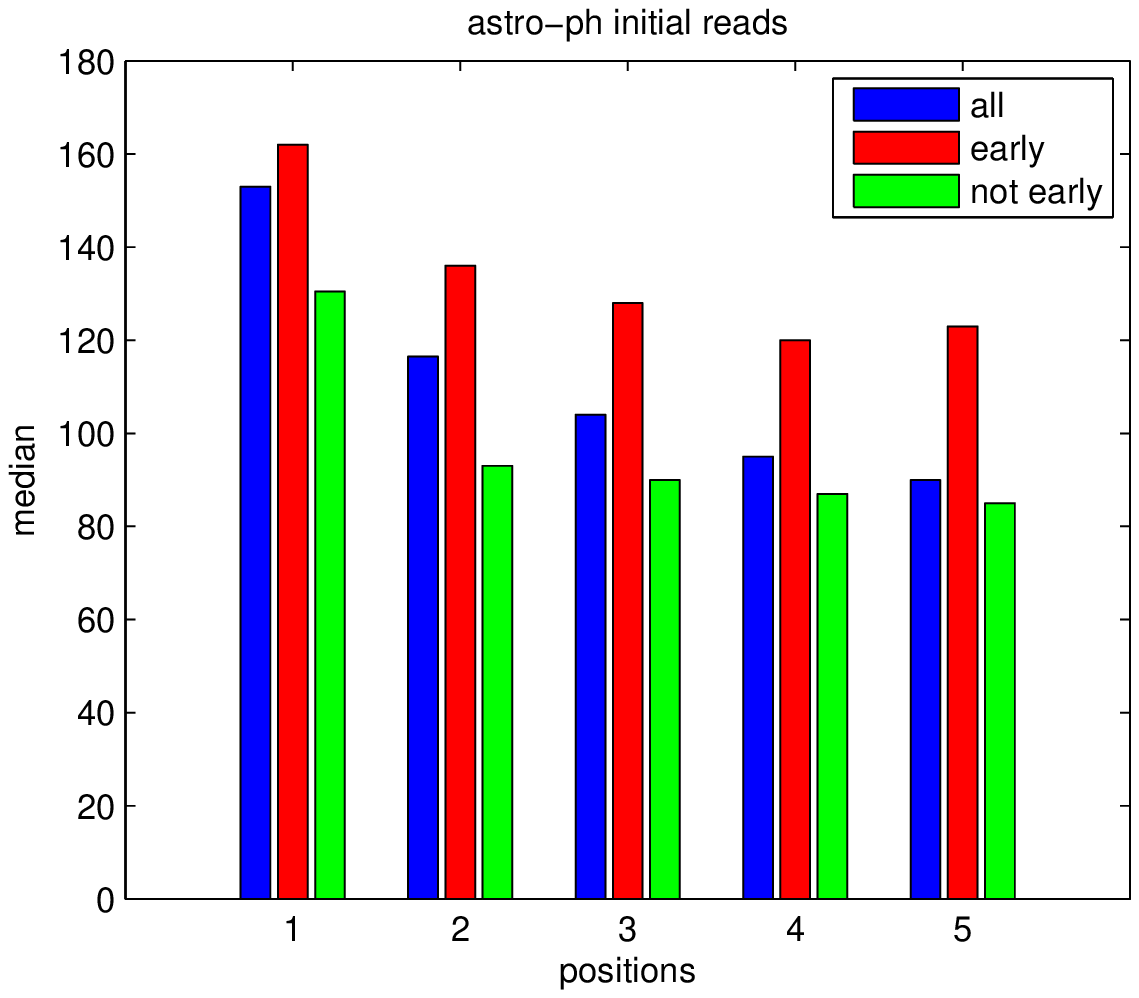}
\centerline{(a)\hskip3.2in(b)}
\caption{\small (a) Box plot of total astro-ph downloads in the active period for different positions. Each box extends from the first through the third quartile, and the red line marks the median. The vertical dashed lines extend above and below to the largest and smallest values within 1.5 times the interquartile range from the respective quartile.  The red `+' signs represent ``outlier'' points above this range.
(b) Median total reads for each position, with the red bars isolating the SP effect and the green bars the V effect.}
\label{fig:astrophinitread}
\end{figure}

\begin{figure}[h!]
\hskip-.2in
\includegraphics[scale=0.65]{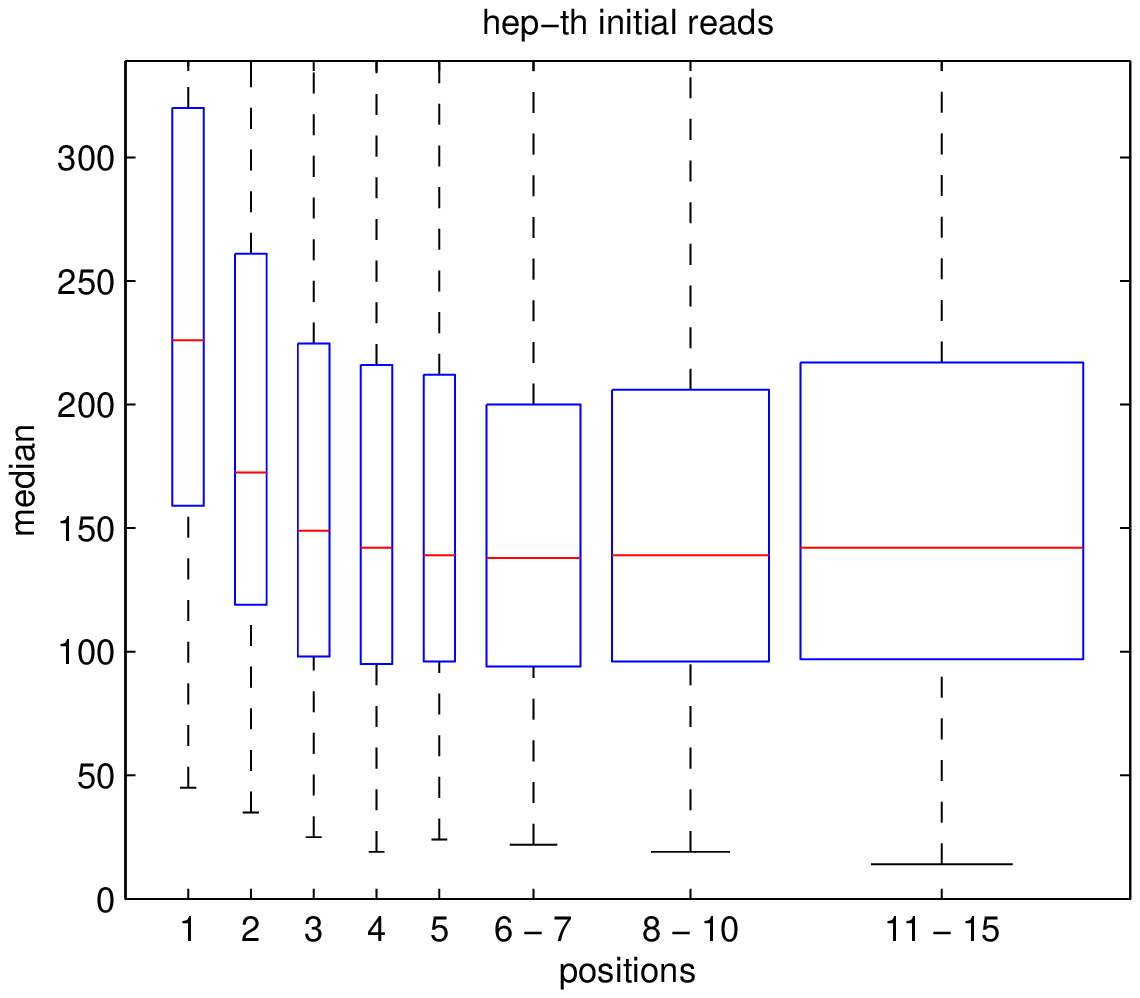}
\quad
\includegraphics[scale=0.65]{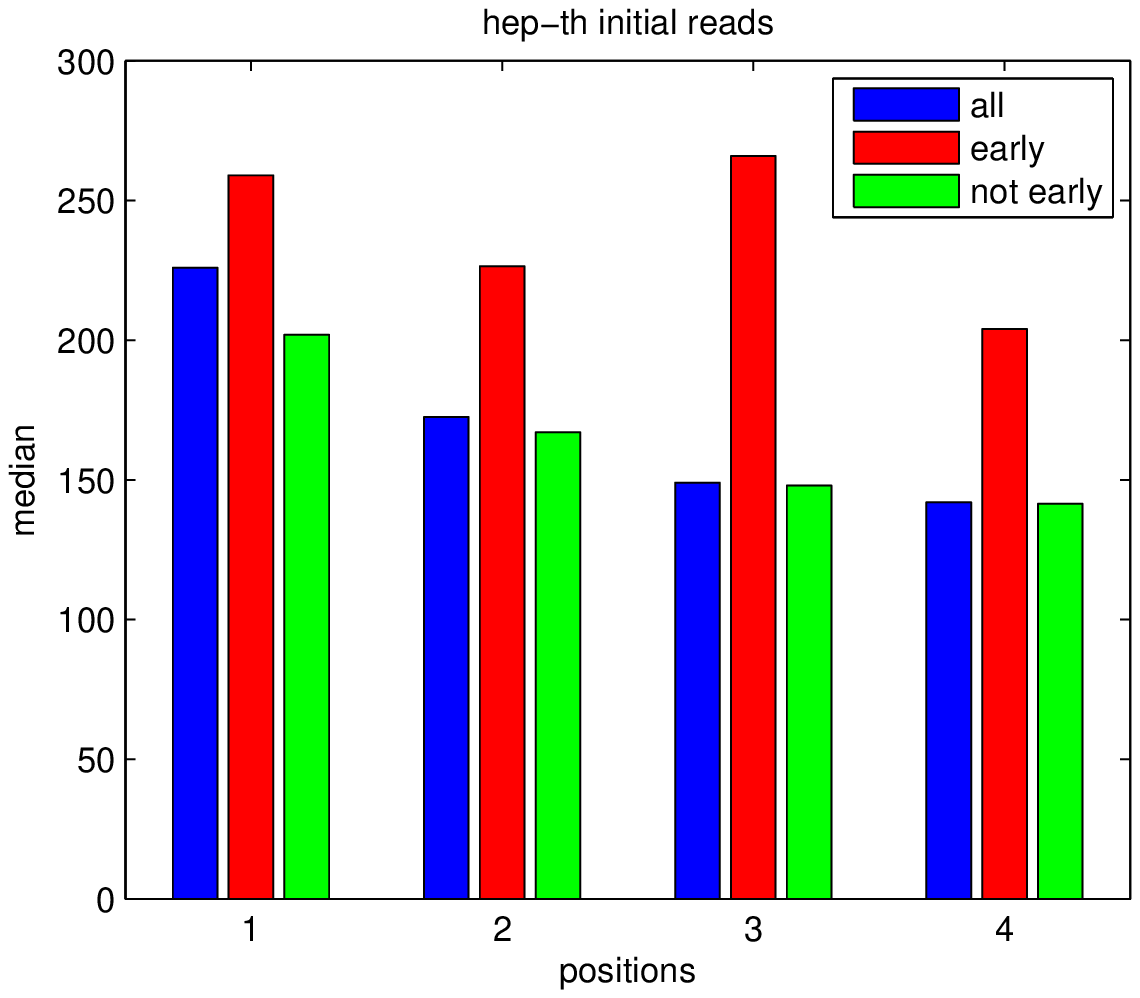}
\centerline{(a)\hskip3in(b)}
\caption{\small (a) Box plot of total hep-th downloads in the active period for different positions, as in fig.~\ref{fig:astrophinitread}a for astro-ph.
(b) Median hep-th reads for each position, as in fig.~\ref{fig:astrophinitread}b for astro-ph.}
\label{fig:hepthinitread}
\end{figure}

\begin{figure}[h!]
\hskip-.2in
\includegraphics[scale=0.65]{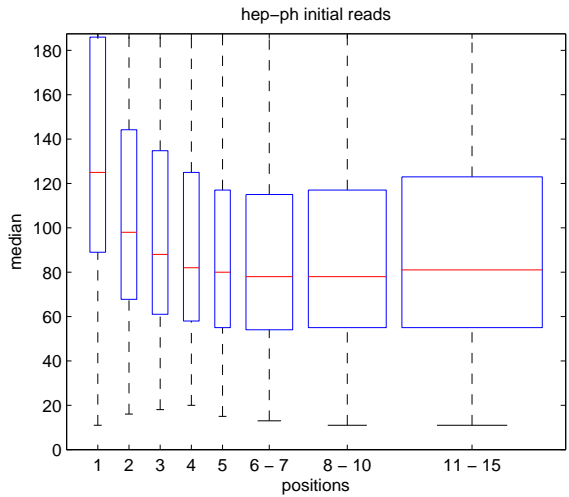}
\quad
\includegraphics[scale=0.65]{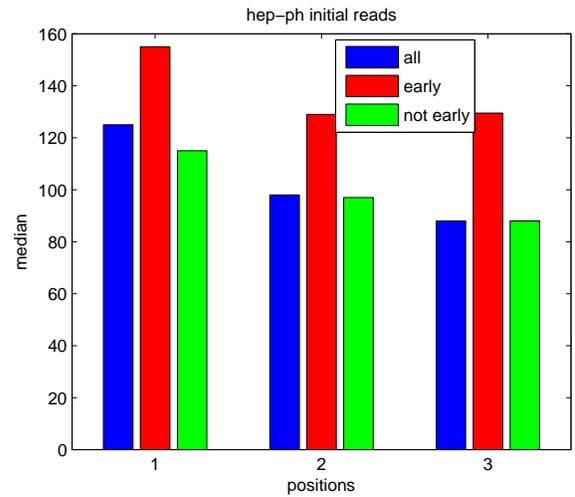}
\centerline{(a)\hskip3in(b)}
\caption{\small (a) Box plot of total hep-ph downloads in the active period for different positions,
as in fig.~\ref{fig:astrophinitread}a for astro-ph and fig.~\ref{fig:hepthinitread}a for hep-th.
(b) Median hep-ph reads for each position, as in fig.~\ref{fig:astrophinitread}b for astro-ph and fig.~\ref{fig:hepthinitread}b for hep-th.}
\label{fig:hepphinitread}
\end{figure}

For astro-ph, we see from fig.~\ref{fig:astrophinitread}a that the number of downloads is higher for the top positions, and the median number declines with position for the early positions. The difference between the first and the second positions is quite striking. 
The differences in medians for the first six positions are statistically significant at the 1\% level. The stochastic dominance of the distributions for different positions is also significant. Position 1 receives roughly twice the median number of initial reads as positions 10--40, indicating a very strong positional effect.
Fig.~\ref{fig:astrophinitread}b shows that the positional effects in readership are dominated by self-promotion: the difference between the red and green bars is significant at the 1\% level for each of the first 5 positions. 
Comparing the green bars, representing ``not early" submissions, with those of fig.~\ref{fig:astrophcitbar}, we also see that the visibility bias is much stronger in the initial popularity of an article than in its long term citations, especially for the first position: the green bars in fig.~\ref{fig:astrophinitread}b show a significant drop from the first position to the next four.\footnote{For comparison with the larger bins mentioned in
subsection~\ref{subsec:DSfV} (footnote after fig.~\ref{fig:astrophcitbar}), astro-ph articles in positions 1--6 received a median of 105 downloads, 44\% higher than the median of 73 for those in positions 10--40.  NE astro-ph articles in positions 1--6 
received a median of 88 downloads, still 20\% higher than for those in positions 10--40.}

For the relation between article position and initial downloads for hep-th and hep-ph, figs.~\ref{fig:hepthinitread}a,\ref{fig:hepphinitread}a show a strong initial download advantage for the first two positions,
and the green bars in figs.~\ref{fig:hepthinitread}b,\ref{fig:hepphinitread}b indicate a strong visibility effect for them. We confirm that the visibility effect can play a strong role in the number of early reads even in the smaller hep-th and hep-ph announcements. 

As pointed out earlier, article position is a one-time artifact of the initial announcement, persisting only for a single day. It is very difficult, if not impossible, to imagine any positional effect on citations in the absence of even stronger positional effects on initial reads. The above initial readership data for astro-ph, hep-th, and hep-ph provide a consistent underpinning for the citation results of the previous section, and are certainly consistent with some form of causal relationship.

\subsection{Correlating Citation with Readership Features}
\label{subsec:citread}

The download data was also analyzed to discover the extent to which article readership predicts citations, and in principle gives some initial measure of article quality. Obvious features that could potentially be correlated with citations are the total downloads, total downloads in the active period, and total number of days getting some downloads. 
Articles whose initial active period is much shorter than average (e.g., 3 days rather than 10) do tend to get somewhat fewer citations in the long run, as would be expected for lower quality articles, rapidly identified as such by discerning readers. In astro-ph, for example, roughly 2.5\% of the articles have 95\% or more of their initial active period downloads during the first 3 days. These receive a median of 4 citations, whereas the remaining articles have a median of 7 citations, a difference statistically significant at the 1\% level. The fraction of active period downloads occurring in the first 3 days could thus be another predictive feature.

It has been observed \cite{Stewart83, Baldi98, vanDalen2001} that the number of citations is positively correlated with the number of authors of an article.
Since articles accumulate citations with time, their age will have some correlation with the number of citations. As discussed earlier, self-promoted early articles receive more citations,  perhaps due to higher quality, and position in the mailing may result in a visibility effect: 
thus whether or not an article is early and its position are important features.

\begin{table}[h]
\centering
\begin{tabular}[c]{| c || c | c | c | c | c | c | c | c |}
\hline
 & E & P & A & AR & F & D & TR & AG\\
\hline
\hline
astro-ph & 0.113 & -0.087 & 0.25 & 0.2753 & 0.069 & 0.326 & 0.328 & 0.086\\
hep-th & 0.07 & 0.013 & 0.256 & 0.4825 & 0.25 & 0.61 & 0.593 & 0.07\\
hep-ph & 0.092 & -0.02 & 0.27 & 0.41& 0.212 & 0.642 & 0.61 & 0.08 \\
\hline
\end{tabular}
\caption{\small Spearman rank correlation between the number of citations with different features: early or not (E), position in mailing (P), number of authors (A), reads in the active period (AR), fraction of active period reads outside the first 3--5 days (F), number of days beyond the active period getting some reads (D), total reads during lifetime (TR), age in days (AG).}
\label{table:allfeaturecorr}
\end{table}

These features are all correlated in some way with the number of citations. We use the citation and readership data for papers submitted between Jan 2002 and Dec 2004 for our analysis. Table~\ref{table:allfeaturecorr} shows the rank correlation between number of citations and the different features individually. The feature most correlated with the ultimate number of citations is the
number of days beyond the active period an article gets some downloads. Steady reads beyond the initial period are thus most predictive of citations, although initial reads are useful as well. 
Reads in this case can even be a consequence of the citations, since citations can lead readers directly to the arXiv site, hence the correlation.\footnote{Whether or not citation or use of bibliographic database leads readers to a journal site after publication or still to the arXiv site depends on how an article is cited, and also on the readership habits of the community, which could differ between the high energy physicists and the astrophysicists. Even the initial period in astro-ph is more likely to share readership with a journal version, since astrophysicists occasionally make arXiv submissions simultaneous with journal acceptance, while high energy physicists tend to make arXiv submissions hot out of the word-processor.}
The total number of reads is also well correlated with citations.

\begin{table}[h]
\centering
\begin{tabular}[c]{| c || c | c | c | c | c | c | c | c |}
\hline
 & E & P & A & AR & F & D & TR & AG\\
\hline
\hline
astro-ph & 0 & 16 & 3 & 69 & 0.15556 & 96 & 185 & 1598\\
hep-th & 0 & 7 & 2 & 137 & 0.17966 & 162 & 346 & 1636\\
hep-ph & 0 & 9 & 2 & 79 & 0.16949 & 118 & 228 & 1630\\
\hline
\end{tabular}
\caption{\small 
Medians of the quantities in  table~\ref{table:allfeaturecorr}.}
\label{table:allfeaturemed}
\end{table}

For completeness, in table~\ref{table:allfeaturemed} we give the medians of the quantities in table~\ref{table:allfeaturecorr}, but emphasize that the details of the distribution reflected in the rank correlation are not well captured by an aggregate quantity like the median. In addition, many of the medians are intrinsically unilluminating. For example, whether or not an article is early (E) is a binary feature, and the median is 0 since more than half the articles are not early.  The median position (P) will be very close to half of the average mailing length since each of the positions has the same number of articles up to that length.
The median age (AG)  is constrained to be roughly 5 years for articles that range from 3.5--6.5 years old, and 
the median reads in the active period (AR) have already been  given in figs.~\ref{fig:astrophinitread}--\ref{fig:hepphinitread}.
Apart from the small fraction of articles that lose readership very quickly, the distribution of the fraction of active period reads after the first 3--5 days (F) will not differ substantially from the overall pattern of exponential falloff in readership.

Is there a meaningful way to harness the combined predictive capacity of the above features? The next logical step beyond correlation is to use regression. In addition to the above features, we have used the daily number of downloads for each of the first 100 days  since the initial period is of much interest, and used the Support Vector Machine implementation SVM$^\text{light}$\footnote{http://svmlight.joachims.org/} \cite{Thorsten99}, a modern supervised machine learning tool (see Appendix~\ref{sec:appSVM}), to predict citations. The methodology involved normalizing every feature by the 95$^{\rm th}$ percentile of its set of values, to avoid convergence problems due to features having values that differ by several orders of magnitude. Since features like the initial and total reads have heavy-tailed distributions, norms like 1-norm,
2-norm or the $\infty$-norm would be dominated by the few large values, and hence normalization by any of these norms would result in setting the small values effectively to zero.

After normalization, the data set was randomly split into five equal parts.
Then we ran $\text{SVM}^\text{light}$ in its regression mode \cite{Smola2003} (with the default linear kernel) five times, using in turn each of the five parts as test set, and the remaining 80\% as training set in each case. This is the standard 5-fold cross-validation procedure to ensure no overfitting of the data. For every run, the predicted citations were compared against the true citations to compute the predictive accuracy. Once again, since citations follow power law distributions, it is preferable to compare the \emph{ranking} of the articles by the predicted citations and the \emph{ranking} produced by the true citations, rather than comparing the actual magnitudes. This was done by computing the Spearman rank correlation coefficient between the predicted citations and the true citations, with the numbers then averaged for the 5 runs.

\begin{table}[h]
\centering
\begin{tabular}{|c || c | c | c |}
\hline
& astro-ph & hep-th & hep-ph \\
\hline
\hline
Average & 0.3930 & 0.5998 & 0.6326 \\
Standard Deviation & 0.0211 & 0.0074 & 0.0168 \\
\hline
\end{tabular}
\caption{\small Spearman rank correlation coefficient between the actual citations and the citations predicted by the SVM regression.}
\label{table:svmresults}
\end{table}

Table~\ref{table:svmresults} shows the extent to which regression was successful in ranking the articles. For hep-th and hep-ph the correlation is indeed quite high. For astro-ph the correlation is smaller, but still substantial.
One possible explanation for this smaller correlation is that astro-ph citations more frequently lead to readership of the journal version, and are not captured by arXiv readership data as well as are citations in the hep-th and hep-ph literatures, whose readers are by habit more likely to consult the version resident on the arXiv server.
To assess this possibility, we folded in data, kindly provided by ADS, giving the number 
of full text downloads directed to the publishers via ADS (rather than to arXiv). This number is strongly correlated  with the number of citations (roughly Spearman 0.5 for articles eventually published in a journal).
Used as an additional feature in our SVM setting, the rank correlation in 
table~\ref{table:svmresults} shifts to 0.7 for astro-ph, now comparable to and even slightly higher than the hep-th and hep-ph correlations. 

\begin{table}[h]
\centering
\begin{tabular}{|c || c | c | c |}
\hline
& astro-ph & hep-th & hep-ph \\
\hline
\hline
Average & 0.3869 & 0.577 & 0.5812 \\
Standard Deviation & 0.0075 & 0.0214 & 0.0200 \\
\hline
\end{tabular}
\caption{\small Spearman rank correlation coefficient between the actual citations and the citations predicted by the SVM regression, but without using the total reads and the long-term fraction of days receiving downloads.}
\label{table:svmresultsmodified}
\end{table}

As noted earlier,  reads beyond the initial period, characterized both by the number of days beyond the active period when a paper gets some reads, and by the total number of reads, are most strongly correlated with citations.
These two features are not necessarily {\em predictive}, however, since later reads at the arXiv site can result in future citations but can also result from citations, either directly or indirectly due to increased interest in an article.
Table~\ref{table:svmresultsmodified} shows the results of removing these two features and again running the SVM regression again with a 5-fold cross-validation.
The correlation weakens slightly, as would be expected, but the early number of reads remains  highly predictive of the long term citation behavior.\footnote{Another highly predictive feature that we did not analyze in detail here is the time to the first citation.}

\subsection{Discussion}

There is no direct analog in other on-line resources for the positional effects on readership considered in sec.~\ref{subsec:poseff}, both due to the nature of the arXiv daily announcements and the central notification role arXiv plays for entire research communities. We've seen that visibility plays a strong unintentional role as a recommender.
The readership effects of the top few positions can be understood in terms of a stochastic decay-of-attention model, in which there is some probability of distraction at each entry, either by pausing to read the associated article full text, or by some external event. The reader either never returns to the original window to read the rest of the list, or having already spent time looking at full text becomes less likely to retrieve later full texts for perusal.  
The difficulty in eliminating such effects provides an additional rationale for offering personalization services  to readers:  when different readers view customized announcements ordered according to their individual preferences,  the artifactual visibility biases of a single global list no longer play a dominant resonant role for the full research community.

The overall correlation we have found between citation and various readership features in table~\ref{table:allfeaturecorr} confirms in a modern electronic context the primary intellectual role played by citation. Rather than playing some symbolic or primarily social role, or thoughtlessly propagated without consultation of sources, citations both appear clearly as a consequence of readership, and lead to further readership. 
The relation found here between readership and later citations amplifies the results of previous studies
\cite{Perneger2004,Moed2004,Brody2005,nature08} 
on the highly predictive role played by early readership.
It is thus tempting to try to incorporate early readership and other newly available measures of popularity, such as blog commentary or other `Web 2.0' commentary mechanisms, into some form of early guide to readers;
and later in an article's lifetime into some more generalized impact metric,  incorporating citations as well.
On the other hand, we've documented here that accidental forms of visibility can drive early readership, with consequent early citation potentially initiating a feedback loop to more readership and citation, ultimately leaving measurable and significant traces in the citation record.
Thus while citations are not primarily used for social purposes, they may nonetheless be subject to indirect influences
familiar from studies of social networking effects \cite{Watts2006}, and thereby not provide an impact metric with the desired objectivity.

Other early activity measures correlated with long-term popularity have recently been considered for on-line sites such as YouTube and Digg \cite{Hub08}, where the effect of early feedback mechanisms is found to be even more pronounced. There are many areas of superficial similarity between on-line scholarship sites on the one hand, and news and commerce sites on the other, but in the context of the results presented here it is important to recall their very different motivations for recommender mechanisms.
An on-line newsite that draws the attention of readers to popular articles increases the number of article reads and hence chances to bring advertisements in front of readers. An on-line commerce site that successfully recommends other popular items increases its number of products ordered and gross revenues. By contrast, a scholarly site that focuses attention on  a smaller number of articles, either intentionally or otherwise, could do an inadvertent disservice to both its authors and readers.

\bigskip\bigskip
\noindent{\bf Acknowledgements:}
We are grateful to Bill Arms, Philip Davis, and Michael Kurtz for comments on a draft manuscript, and thank Philip Davis for pointing us to related prior work.

\appendix
\section{Power Law Fitting}
\label{sec:appPLF}

To fit data to a power law, often the method of maximum likelihood estimation is used to compute the power law exponent, followed by a least squares fit of a straight line with the computed slope in a log-log plot.

For a power law distribution with $p(x) \propto x^{-\alpha}$, the cumulative distribution $F(X > x) \propto x^{-(\alpha-1)}$, is also a power law.
The cumulative distribution $F(X > x)$ is smoother so is customarily used for power law fitting, and is often plotted as a Rank-Frequency (RF) plot \cite{Newman04}. Swapping the axes of an RF plot gives the Zipf plot, which follows a power law behavior with the inverse of the RF exponent.

\cite{Dietrich07} gives the Zipf plots of citations for the top 10 positions and the remaining positions, binned appropriately. These curves give the cumulative distribution function $F(X > x)$ for different positions, and are useful in comparing two distributions for stochastic dominance.
A cumulative distribution $F(X > x)$ is said to \emph{stochastically dominate (first order)} \cite{Bawa75} another cumulative distribution $G(X > x)$ iff for all $x$ we have
\[
F(X > x) \ge G(X > x) \ .
\]
In risk analysis, it is always safer to gamble according to the dominating distribution,
since it is expected to produce higher values.
 If one RF curve is always above another, then there is stochastic dominance, although the statistical significance of the dominance needs to be verified.

In \cite{Dietrich07}, the citation distribution of the top position is found to be higher than the lower positions.
The power law exponent of the Zipf plot in \cite{Dietrich07} is $\beta = 0.48$, in accord with \cite{Redner98}. The power law exponent of the citation distribution is thus $\alpha = 1+\frac{1}{\beta} = 3.0833$. At this value of $\alpha$, the mean \cite{Newman04} citation is
\[
\langle x \rangle = \frac{\alpha-1}{\alpha-2}x_\text{min} \ .
\]
If there is an upper limit, $x_\text{max}$, then the mean becomes
\[
 \langle x \rangle = \frac{\alpha-1}{\alpha-2}x_\text{min}\left[1 - {\left(\frac{x_\text{min}}{x_\text{max}}\right)}^{\alpha-2}\right]\ . 
 \]

\def\elfo{10\hbox{\small--}40}
To restrict to the region where the power law is valid, the small and large rank regions of the Zipf plots
are excluded in \cite{Dietrich07}, introducing (a) a normalization bias, and (b) a potential bias of eliminating a large fraction of the data; as we now describe:\\
\indent $\bullet$
(a) Given two curves, say citations corresponding to position 1,
 and to positions 10--40,
 the restriction to the power law region introduces cutoffs  $x^1_\text{min} > x^{\elfo}_\text{min}$  and $x^1_\text{max} > x^{\elfo}_\text{max}$, where
\[
\log x^1_\text{min} - \log x^{\elfo}_\text{min} = \log x^1_\text{max} - \log x^{\elfo}_\text{max}
\]
(since log-log plots of two parallel straight lines are equidistant at the endpoints). This gives
\[
\begin{array}{r c l}
\frac{x^1_\text{min}}{x^1_\text{max}} &=& \frac{x^{\elfo}_\text{min}}{x^{\elfo}_\text{max}}\\
\implies 1 - {\left(\frac{x^1_\text{min}}{x^1_\text{max}}\right)}^{\alpha-2} &=& 1 - {\left(\frac{x^{\elfo}_\text{min}}{x^{\elfo}_\text{max}}\right)}^{\alpha-2}\\
\implies\qquad \frac{\langle x^1 \rangle}{\langle x^{\elfo} \rangle} &=& \frac{x^1_\text{min}}{x^{\elfo}_\text{min}}\ .\\
\end{array}
\]
The cut-off in  \cite{Dietrich07} was such that $\frac{x^1_\text{min}}{x^{\elfo}_\text{min}} \approx 2$, so it is not clear whether the factor of 2 advantage in the average was due to the cut-off having given $\langle x^1 \rangle$ the benefit of higher $x_\text{min}$.\\
\indent $\bullet$
(b) Our analysis of the same data gives a median citation for position 1 of 10, and for positions 10--40 of 4. 
A large lower cutoff will thus ignore a large fraction of the data.
Ref.~\cite{Dietrich07} used $x^1_\text{min} \approx 50$, whereas the $75^\text{th}$ percentile of the citations for position 1 is 22. This means at least $\frac{3}{4}$ of the data was ignored to compute the aggregate values.

\section{Statistical Significance}
\label{sec:appSS}

To test the statistical significance of the difference in median citations, we use the Mann-Whitney U (MWU) test, with the null hypothesis that the medians are equal, and the two-sided alternative that the medians are not equal, at 1\% significance level. Table~\ref{table:astromwucitbox} shows that for astro-ph the medians of the top 5 positions are significantly different from the medians of the positions 10 and beyond.

\begin{table}[h]
\centering
\begin{tabular}[c]{| c || c |}
\hline
Position from & Position onwards \\
\hline
\hline
1 & 2 \\
2 & 5 \\
3 & 5 \\
4 & 7 \\
5 & 11 \\
6 & 11 \\
\hline
\end{tabular}
\caption{\small \emph{Mann-Whitney U test for astro-ph}. Left column is the position whose median we are assessing for significant difference  (1\% significance level) with a two-sided alternative (either median the greater).
The right column is the position whose median (and that of positions beyond) is significantly different from the corresponding position on the left column. For example the median number of citations for position 2 is greater than that of positions 5 and beyond, at 1\% significance level.}
\label{table:astromwucitbox}
\end{table}

\begin{table}[h]
\centering
\begin{tabular}[c]{| c || c |}
\hline
Position from & Position onwards \\
\hline
\hline
1 & 4 \\
2 & 5 \\
3 & 6 \\
4 & 7 \\
5 & 11 \\
\hline
\end{tabular}
\caption{\small \emph{Kolmogorov-Smirnov test for astro-ph}. Left column is the position whose distribution we are assessing for stochastic domination (1\% significance level) with a one-sided alternative. The right column is the position whose distribution (and the positions beyond) is stochastically dominated by the corresponding position on the left column. For example the median number of citations for position 2 is greater than that of positions 5 and beyond, at all levels, at 1\% significance level.}
\label{table:astrokscitbox}
\end{table}

A significant difference in median does not necessarily mean a distribution is better at all levels. To test stochastic domination, we used the Kolmogorov-Smirnov (KS) test with the null hypothesis that the two distributions are the same, and the one-sided alternative
that the first distribution dominates the second, at 1\% significance level. Table~\ref{table:astrokscitbox} shows that for astro-ph the first 5 positions are indeed better than all other positions, for all values.

\section{SVM Regression}
\label{sec:appSVM}

SVM regression \cite{Smola2003} is different from the standard regression task in two ways. Firstly, SVM uses the 
$\varepsilon$-insensitive loss function where for individual sample points only an error of greater than $\varepsilon$ counts as ``error'', and the total error is the sum of the samplewise errors. Secondly, the minimization function is a combination of the $\varepsilon$-insensitive loss function as well as the squared norm of the vector of regression coefficients. The tradeoff between this norm and the loss function is controlled by a parameter $C$. The algorithm takes both $\varepsilon$ and $C$ as parameters, and setting 
small $\varepsilon$ and large $C$ gives a form of least squares result. SVM regression uses state of the art constrained optimization techniques to find a solution. Its real power, however, is the ease with which nonlinearity can be incorporated by higher order kernels. The efficiency and accuracy of this approach has already been established firmly in the realm of machine learning through numerous principled applications.

To explore the predictive capacity of readership and other features, we treated it as a standard supervised prediction task in machine learning. Some past attempts to correlate citations with article and author features \cite{Stewart83, Baldi98, vanDalen2001} used samples that were several orders of magnitude smaller and hence allowed manual extraction of features.\footnote{While manual extraction of features is not as feasible on the larger datasets currently in use, modern text-mining tools together with the increased availability of the full-texts in digital form should ultimately permit  automated extraction of a comparable set of features.}
In such a setting regression is used for the entire dataset and the total error is reported. A potential problem with this approach is that it may simply validate the regression model used, rather than result in learning and prediction. Use of the full dataset may also be vulnerable to overfitting through extraneous features. In machine learning, the standard approach is to cross-validate through random training and test splits of the data, and report the average accuracy on the test sets. This puts less emphasis on human verification of the model being learned, especially when higher order kernels are used. 

\newpage
\bibliographystyle{abbrv}
\bibliography{positionaleffects}

\begin{thebibliography}{10}

%

\bibitem[Baldi, 1998]{Baldi98}
Baldi, S., (1998).
\newblock Normative versus Social Constructivist Processes in the Allocation of
  Citations: A Network-Analytic Model.
\newblock American Sociological Review, 63, 829--846.

\bibitem[Bawa, 1975]{Bawa75}
Bawa, V. S. (1975).
\newblock Optimal Rules for Ordering Uncertain Prospects.
\newblock Journal of Financial Economics, 2, 95--121.

\bibitem[Brody et al., 2006]{Brody2005}
Brody, T., Harnad, S., \& Carr, L. (2006).
\newblock Earlier Web Usage Statistics as Predictors of Later Citation Impact.
\newblock JASIST, 57, 1060--1072.

\bibitem[Davis \& Fromerth, 2007]{DavFro07}
Davis, P. M. and Fromerth, M. J. (2007).
\newblock Does the arXiv lead to higher citations and reduced publisher
downloads for mathematics articles?
\newblock JASIST, 71, 203--215.

\bibitem[Davis et al., 2008]{Davis08}
Davis, P. M., Lewenstein, B. V., Simon, D. H., Booth, J. G., Connolly, M. J. L. (2008).
\newblock Open access publishing, article downloads, and citations: randomised
controlled trial.
\newblock BMJ 337, a568.
\newblock\url{http://dx.doi.org/10.1136/bmj.a568}

\bibitem[Dietrich, 2008a]{Dietrich07}
Dietrich, J. P. (2008a).
\newblock The importance of being first: Position dependent citation rates on
  arxiv:astro-ph.
\newblock PASP 120, 224--228.
\hfill\break\newblock \url{http://arxiv.org/abs/0712.1037}

\bibitem[Dietrich, 2008b]{Dietrich08}
Dietrich, J.P. (2008b).
\newblock Disentangling visibility and self-promotion bias in the
  arxiv:astro-ph positional citation effect.
\newblock PASP 120, 801--804.
\hfill\break\newblock \url{http://arxiv.org/abs/0805.0307}

\bibitem[Fortunato et al., 2006]{Fort06}
Fortunato, S., Flammini, A., Menczer, F.,  Vespignani, A. (2006).
\newblock Topical interests and the mitigation of search engine bias.
\newblock PNAS, 103, 12684--12689.
\newblock\url{http://dx.doi.org/10.1073/pnas.0605525103}

\bibitem[Gibbons, 1997]{Gibbons97}
Gibbons, J. D. (1997).
\newblock {\em Nonparametric Methods for Quantitative Analysis}.
\newblock American Sciences Press ISBN-13: 978-0935950373.

\bibitem[Ginsparg, 2007]{Ginsparg07}
Ginsparg, P. (2007).
\newblock Next-Generation Implications of Open Access.
\newblock CTWatch Quarterly, 3(3), August 2007.

\bibitem[Granka et al., 2004]{Granka04}
Granka, L., Joachims, T., and Gay, G. (2004).
\newblock Eye-Tracking Analysis of User Behavior in WWW Search.
\newblock In {\em Proceedings of the
28th Annual ACM Conference on Research and Development in Information and
Retrieval}. (SIGIR '04). Sheffield, UK.
\hfil\break\newblock \url{http://www.cs.cornell.edu/People/tj/publications/granka_etal_04a.pdf}

\bibitem[Joachims, 1999]{Thorsten99}
Joachims, T. (1999).
\newblock Making large-Scale SVM Learning Practical.
\newblock in Sch\"olkopf, B., Burges, C., \& Smola, A. (ed.),
{\em Advances in Kernel Methods - Support Vector Learning},  MIT-Press.
\hfil\break\newblock \url{http://www.joachims.org/publications/joachims_99a.pdf}


\bibitem[Kurtz et al., 2005a]{Kurtz04}
Kurtz, M.J., Eichhorn, G., Accomazzi, A., Grant, C.,                 
Demleitner, M., Murray, S., Martimbeau, N., Elwell, B. (2004).
\newblock The bibliometric properties of article readership information.
\newblock JASIST, 56, 111--128.
\newblock\url{http://dx.doi.org/10.1002/asi.20096}

\bibitem[Kurtz et al., 2005b]{Kurtz05}
Kurtz, M.J., Eichhorn, G., Accomazzi, A., Grant, C., Demleitner, M., Henneken, E., Murray, S. (2005).
\newblock The effect of use and access on citations.
\newblock Inform Process Manag, 41, 1395--1402.

\bibitem[Moed, 2005]{Moed2004}
Moed, H. F. (2005).
\newblock Statistical Relationships Between Downloads and Citations at the
  Level of Individual Documents Within a Single Journal.
\newblock JASIST, 56, 1088--1097.

\bibitem[Moed, 2007]{Moed07}
Moed, H. F. (2007).
\newblock The effect of ``Open Access'' upon citation impact: An analysis of ArXiv's Condensed Matter Section.
\newblock JASIST, 58, 2047--2054.

\bibitem[Neurosci Editor, 2008]{nature08}
Editorial (2008).
\newblock Deciphering citation statistics.
\newblock Nature Neuroscience 11, 619.
\newblock\url{http://dx.doi.org/10.1038/nn0608-619}

\bibitem[Newman, 2005]{Newman04}
Newman, M. E. J. (2005).
\newblock Power laws, Pareto distributions and Zipf's law.
\newblock Contemporary Physics 46, 323--351.
\newblock \url{http://arxiv.org/abs/cond-mat/0412004}

\bibitem[Perneger, 2004]{Perneger2004}
Perneger, T. V. (2004).
\newblock Relation between online ''hit counts'' and subsequent citations:
  prospective study of research papers in the BMJ.
\newblock BMJ, 329, 546--547.

\bibitem[Phillips et al., 1991]{Phillips91}
Phillips, D. P., Kanter, E. J., Bednarczyk, B., Tastad, P. L. (1991).
\newblock Importance of the lay press in the transmission of medical knowledge to the scientific community.
\newblock NEJM, 325(16), 1180--1183.

\bibitem[Redner, 1998]{Redner98}
Redner, S. (1998).
\newblock How Popular is your Paper? An Empirical Study of the Citation
  Distribution.
\newblock Eur.Phys.J. B4, 131--134.
\hfil\break\newblock \url{http://arxiv.org/abs/cond-mat/9804163}

\bibitem[Salganik, et al., 2006]{Watts2006}
Salganik, M. J., Dodds, P. S., \& Watts, D. J. (2006).
\newblock Experimental Study of Inequality and Unpredictability in an
  Artificial Cultural Market.
\newblock Science, 311, 854--856.
\newblock\url{http://dx.doi.org/10.1126/science.1121066}

\bibitem[Smola \& Sch\"olkopf, 2004]{Smola2003}
Smola, A. J. \& Sch\"olkopf, B. (2004).
\newblock A tutorial on support vector regression.
\newblock Statistics and Computing, 14, 199--222.
\hfil\break\newblock \url{http://alex.smola.org/papers/2003/SmoSch03b.pdf}

\bibitem[Stewart, 1983]{Stewart83}
Stewart, J. A. (1983).
\newblock Achievement and Ascriptive Processes in the Recognition of Scientific
  Articles.
\newblock Social Forces, 62, 166--189.

\bibitem[Szabo \& Huberman, 2008]{Hub08}
Szabo, G., Huberman, B. A. (2008).
\newblock Predicting the popularity of online content.
\newblock\url{http://arxiv.org/abs/0811.0405}

\bibitem[van Dalen \& Henkins, 2001]{vanDalen2001}
van Dalen, H. P. \& Henkens, K. (2001).
\newblock What makes a scientific article influential? The case of
  demographers.
\newblock Scientometrics, 50, 455--482.

\end{thebibliography}

\end{document}